\documentclass[smallextended]{svjour3}
\usepackage{url}
\usepackage{graphicx}
\usepackage{amssymb}
\usepackage{natbib}
\usepackage{aas_macros}
\usepackage{soul}
\usepackage{tcolorbox}
\begin{document}

\title{Exploring the link between star and planet formation with Ariel}



\author{Diego Turrini, Claudio Codella, Camilla Danielski, Davide Fedele, Sergio Fonte, Antonio Garufi, Mario Giuseppe Guarcello, Ravit Helled, Masahiro Ikoma, Mihkel Kama, Tadahiro Kimura, J.~M.~Diederik Kruijssen, Jesus Maldonado, Yamila Miguel, Sergio Molinari, Athanasia Nikolaou, Fabrizio Oliva, Olja Pani\'c, Marco Pignatari, Linda Podio, Hans Rickman, Eugenio Schisano, Sho Shibata, Allona Vazan, Paulina Wolkenberg}

\authorrunning{Turrini et al.}

\institute{ 
            Turrini D.\at 
            INAF - Osservatorio Astrofisico di Torino, Via Osservatorio 20, I-10025, Pino Torinese, Italy\\ 
            Institute of Space Astrophysics and Planetology INAF-IAPS, Via Fosso del Cavaliere 100, I-00133, Rome, Italy, \email{diego.turrini@inaf.it}
            \and            
            Fonte S., Schisano E., Molinari S., Oliva F., Wolkenberg P. \at Institute of Space Astrophysics and Planetology INAF-IAPS, Via Fosso del Cavaliere 100, I-00133, Rome, Italy
            \and
            Fedele D. \at 
            INAF - Osservatorio Astrofisico di Torino, Via Osservatorio 20, I-10025, Pino Torinese, Italy\\
            INAF - Osservatorio Astrofisico di Arcetri, Largo E. Fermi 5, I-50127 Firenze, Italy
            \and
            Codella C., Garufi A., Podio L. \at INAF - Osservatorio Astrofisico di Arcetri, Largo E. Fermi 5, 50127 Firenze, Italy
            \and
            Danielski C. \at
            Instituto de Astrof\'isica de Andaluc\'ia (IAA-CSIC), Glorieta de la Astronom\'ia s/n, 18008 Granada, Spain
            \and
            Guarcello M.~G., Maldonado J. \at
            INAF - Osservatorio Astronomico di Palermo, Piazza del Parlamento 1, I-90134, Palermo, Italy
            \and
            Helled R. \at
            Institute for Computational Science, Center for Theoretical Astrophysics \& Cosmology University of Zurich, CH-8057 Zurich, Switzerland
            \and
            Ikoma M., Kimura T., Shibata S. \at
            Department of Earth and Planetary Science, The University of Tokyo, 7-3-1 Hongo, Bunkyo-ku, Tokyo 113-0033, Japan
            \and
            Kama M. \at
            Department of Physics and Astronomy, University College London, London, WC1E 6BT, UK\\
            Tartu Observatory, University of Tartu, Observatooriumi 1, 61602, T\~{o}ravere, Estonia
            \and
            Kruijssen J.~M.~D. \at
            Astronomisches Rechen-Institut, Zentrum f\"{u}r Astronomie der Universit\"{a}t Heidelberg, M\"{o}nchhofstra\ss e 12-14, 69120 Heidelberg, Germany
            \and
            Miguel Y. \at
            Leiden Observatory, Leiden University, Niels Bohrweg 2, 2333CA Leiden, The Netherlands\\
            SRON - Netherlands Institute for Space Research, Sorbonnelaan 2, NL-3584 CA Utrecht, The Netherlands
            \and
            Pani\'c O. \at
            School of Physics and Astronomy, E.~C.~Stoner Building, University of Leeds, Leeds LS2 9JT, United Kingdom
            \and
            Rickman H. \at
            Centrum Bada\'n Kosmicznykh Polskiej Akademii Nauk (CBK PAN), Bartycka 18A, 00-716 Warszawa, Poland
            \and
            Pignatari M. \at
            E.A. Milne Centre for Astrophysics, Department of Physics \& Mathematics, University of Hull, HU6 7RX, United Kingdom \\
            Konkoly Observatory, Research Centre for Astronomy and Earth Sciences, Hungarian Academy of Sciences, Konkoly Thege Miklos ut 15-17, H-1121 Budapest, Hungary \\
            Joint Institute for Nuclear Astrophysics - Center for the Evolution of the Elements, USA \\
            NuGrid Collaboration, \protect\url{www.nugridstars.org}
            \and
            Nikolaou A. \at
            Sapienza University of Rome, Piazzale Aldo Moro 2, 00185, Italy\\
            European Space Agency, ESRIN, ESA $\Phi$-lab, Largo Galileo Galilei 1, 00044 Frascati, Italy
            \and
            Vazan A. \at
            Department of Natural Sciences and Astrophysics Research Center of the Open university (ARCO), The Open University of Israel, 4353701 Raanana, Israel. 
            }

\maketitle

\begin{abstract}
The goal of the Ariel space mission is to observe a large and diversified population of transiting planets around a range of host star types to collect information on their atmospheric composition. The planetary bulk and atmospheric compositions bear the marks of the way the planets formed: Ariel's observations will therefore provide an unprecedented wealth of data to advance our understanding of planet formation in our Galaxy. A number of environmental and evolutionary factors, however, can affect the final atmospheric composition. Here we provide a concise overview of which factors and effects of the star and planet formation processes can shape the atmospheric compositions that will be observed by Ariel, and highlight how Ariel's characteristics make this mission optimally suited to address this very complex problem.
\end{abstract}



\section{Introduction}\label{sec:introduction}

The study of the initial stages of the life of planetary systems, when planets are forming within the gaseous embrace of protoplanetary discs, has been undergoing a transformation in recent years. The improved resolution of observational facilities is allowing us to directly observe, for the first time, the gaps and rings in the gas and dust of protoplanetary discs that were the theoretically predicted signatures of the appearance of giant planets. These observations are being accompanied by improvements in the compositional characterisation of the discs themselves, allowing the first direct comparisons between the volatile budgets in protostellar objects and in the comets of our Solar System.

These advances have proceeded in parallel with the continuous growth of the known population of extrasolar planets, whose current size exceeds 4000 members and is allowing for population studies at the level of individual planets, of planetary systems as a whole, and of the link between stellar and planetary characteristics. The overall picture emerging from all these fields of study, while still incomplete, is nevertheless clearly indicating how the characteristics of each individual planet are uniquely sculpted by those of the environment in which it forms, in turn set by the star and its own formation process.

Ariel \citep{tinetti+2018}, the M4 mission of the European Space Agency deemed for launch in \textcolor{black}{2029}, will characterise the composition of hundreds of exoplanetary atmospheres, proving us with an unprecedentedly large and diversified observational sample \citep{tinetti+2018,zingales+2018,turrini+2018,edwards+2019}. Ariel's observations will further revolutionize our view of the formation and evolution of both individual planets and planetary systems by systematically introducing a new dimension, their atmospheric composition, in the study of these subjects. 

The insight on the link between the star formation process and the compositional build of planets will become an increasingly important piece of the puzzle of unveiling the nature of exoplanets over the coming years. The goal of this paper is therefore twofold. The first part (Sects. \ref{sec:protoplanetary_discs}-\ref{sec:organics}) aims to explore the environmental factors linked to the star formation process and the evolution of protoplanetary discs that can impact the final build of the exoplanets that Ariel will observe. At the beginning of Ariel's observational campaign these factors might represent a source of uncertainty in the interpretation of the atmospheric data (e.g. an unknown composition of the host star). By the end of the nominal mission, however, Ariel's rich and diverse exoplanetary sample will \textcolor{black}{allow} to shed light on the interplay between these environmental factors and the planet formation process.

The second part (Sects. \ref{sec:giant_planets}-\ref{sec:architectures}), which builds upon the science cases devised in the previous phases of the mission \citep{tinetti+2018,turrini+2018}, aims to detail more closely the implications of the planet formation process for Ariel's observations. Specifically, Sect. \ref{sec:giant_planets} will discuss in detail the case of giant planets, which currently represent the bulk of Ariel's observational sample \citep{edwards+2019}, while Sect. \ref{sec:high_density_planets} will move to smaller planetary sizes and masses, discussing the interplay between the capture of the nebular gas and the outgassing from the planetary interior in shaping their atmospheres as primary, secondary or mixed. \textcolor{black}{Finally, Sect. \ref{sec:architectures} will review the information that can be extracted by the architectures of the planetary systems hosting the planets that Ariel will observe to provide dynamical context to the interpretation of Ariel's atmospheric data.} 
Because of the interdisciplinary nature of the discussion throughout this paper, each section and subsection will \textcolor{black}{include} one \textit{coloured box} providing the associated \textit{take-home message}, to help readers to identify and connect the key points for all specific subjects discussed. 

\section{Circumstellar Discs as the Birth Environment of Planets}\label{sec:protoplanetary_discs}

The cores and atmospheres of planets are composed of protostellar dust, ices, and gas that have undergone physical and chemical interactions in planet-forming discs around newborn stars. To interpret the full range of planetary properties that Ariel will reveal, and to contribute to its list of science questions, it is therefore critical to have a firm foundation in this stage. 

The formation of stars of mass up to several M$_\odot$ is accompanied by the emergence of a flattened \textcolor{black}{expanse} of material in Keplerian rotation around the star, called the protoplanetary disc. First signs of protoplanetary discs are present from the early infall stages when the star is accreting significantly \cite[Class 0-I, ][]{Lada1987}. \textcolor{black}{Class 0 is the stage where the protostar is fully embedded in its parent envelope and rapidly gains its mass. In the Class I stage, the star has already accreted most of its mass. In both these stages, the disc is the channel for accretion onto the star from the surrounding envelope, but it only becomes observationally accessible in Class I stage. The most accessible phase, observationally, is the Class II phase where the star has accumulated its mass almost entirely, becomes directly visible and is no longer embedded in its parent envelope, leaving only the protoplanetary disc.}

Across these stages, the disc is the channel for accretion onto the star from the surrounding envelope, and accretion may proceed up to several Myr on the pre-main sequence. Emerging evidence from the ALMA and VLA interferometers \textcolor{black}{indicates that the \emph{average} mass of discs arond Solar-like stars decreases from $10^{-1}$\,M$_{\odot}$ at Class~0 to $10^{-1.5}$\,M$_{\odot}$ and $10^{-2}$\,M$_{\odot}$ at Class~I and II, respectively \cite[e.g.][]{tychoniec+2018}}. In other words, the planet-forming mass reservoir drops from $100$\,M$_{\rm J}$ to $\leq10$\,M$_{\rm J}$ (Jovian masses) when moving from embedded to easily observable discs. \textcolor{black}{While the matter will be discussed in more detail later in the text, it is worth mentioning already here that these disc mass estimates suffer from a number of caveats mainly due to our poor knowledge of the dust opacity and the gas-to-dust mass ratio. The latter is often assumed to be similar to that of the interstellar medium (ISM), whose gas-to-dust mass ratio is estimated being 100:1 \citep{bohlin+1978}.}

Discs extend up to a few hundred au, with several known cases extending over 500~au. The disc mass consists almost entirely of gas, so the hydrostatic pressure opposes the gravitational pull \textcolor{black}{toward the plane of rotation and supports an extended vertical structure. The disc vertical struture is characterized in terms of the scale height, defined as $h_R = H/R$ with $H$ being the height above the disc midplane in au, and R the distance from the star in au. $H$ in turn is defined as the ratio between the local sound speed $c_s$ and orbital angular velocity $\Omega$. Typical values of $h_R$ range between a few $10^{-2}$ to 0.1 depending on the temperature profile, hence the sound speed, of the disc \citep[e.g.][]{dalessio+2001,armitage2009}, though recent evidence shows that gas and small dust grains can reach scale height of 0.15-0.25 at R$>$ 100 au \citep[e.g.][]{avenhaus+2018}. While the gas and small grains can reach such altitudes, large dust grains settle fast onto the disc midplane as shown e.g., by the ALMA survey of edge-on discs \citep{villenave+2020}.} 

At the final stage, Class III, the star has typically already reached the main sequence, and can be surrounded by a debris disc (i.e. Vega-like stars, \citealt{dominik+2000}). A debris disc is an extremely flat disc containing solids only, and rather than a disc it is geometrically better described as one or more rings or belts, analogous to our \textcolor{black}{asteroid and Kuiper belts} \citep{matthews+2014}. With the high sensitivity of ALMA, we are finding that the boundary between protoplanetary and debris discs is blurred \citep{Miley+2018}, with some debris discs containing gas, although much less than typical protoplanetary discs \citep{pericaud+2016}, and protoplanetary discs showing evidence of dust production by collisional mechanisms as debris discs \citep{turrini+2019}.

\textcolor{black}{Two main processes have been proposed as possible pathway for the formation of giant planets, the \textit{core accretion} (also called nucleated instability) scenario and the \textit{disc instability} scenario, with different implications for the disc environment associated with the birth of these planets. We briefly highlight the main characteristics of these two processes in the following, referring the readers to the recent reviews by \cite{helled+2014} and \cite{dangelo+2018} for more in-depth discussions. \\
\indent In the disc instability scenario giant planets form as a result of a local gravitational instability in the circumstellar disc, which leads to the formation of a gravitationally bound object that collapses under its own self-gravity on timescales of the order of a few to a few tens of orbital periods. 
Disc instability 
may happen through Classes 0-I, when the disc is massive and more likely to be unstable. The condition for disc instability is satisfied for low values of the Toomre parameter}
\begin{equation}
G_T = \frac{c_s \Omega}{\pi G \Sigma} < 1
\end{equation}
\textcolor{black}{\noindent with G being the gravitational constant and $\Sigma$ the gas surface density. Thus, for disc instability to occur the disc must be cold and massive, a condition that can be easily satisfied in the outer region of very young discs (roughly beyond a few tens of au). In Class II, and by the time the envelope is gone, it is likely that the disc has also had sufficient time to reach stability although, observationally, it is quite difficult to exclude the possibility that some of the massive discs with spiral structures seen in Class II may be undergoing disc instability. \\
\indent In the core accretion scenario (further discussed in Sect. \ref{sec:giant_planets}) the giant planets first form a planetary core by accumulation of solid material in the inner and denser regions of protoplanetary discs (within the first few tens of au), meanwhile acquiring a more or less extended gaseous envelope by capturing gas from the circumstellar disc. When the mass of this expanded atmosphere becomes comparable with that of the planetary core, the gas becomes gravitationally unstable and triggers a runaway gas infall phase that causes a very rapid mass growth of the planet. The time required for the planetary core to grow and trigger the instability of its extended atmosphere can vary between a few 10$^5$-10$^6$ years, while the runaway gas accretion timescale is an order of magnitude faster, ranging between a few 10$^4$ years and a few 10$^5$ years. 
Due to its longer timescales nucleated instability can operate well into Class II, as the time required for the growth of the planetary core can easily exceeds the age of typical Class I sources.  
} 

Because of the wealth of observationally constrained parameters of discs in Class II phase, the information we possess on discs is mainly related to \textcolor{black}{this class},  
with only a few examples which could qualify as representative of \textcolor{black}{disc instability} conditions.

\subsection{Gas and solids in protoplanetary discs} 

Almost the entire gas mass of the disc is in the form of molecular hydrogen (H$_2$) and helium (He), with the next most abundant molecule being CO, with abundance of $\le$10$^{-4}$ with respect to H. Direct total H$_{2}$ gas mass measurements are not feasible because H$_{2}$ lacks a permanent dipole moment, which makes its rotational emission unobservably weak. The first vibrational level requires $\sim 6000\,$K to excite, so ro-vibrational emission only probes a thin, hot surface layer of the inner disc, same as the fluorescent emission of H$_2$ \citep{thi+2001}. Consequently the gas mass pursuit has focused on species such as CO and HD.

The \textit{H$_{2}$ isotopologue, HD}, has a permanent dipole moment and rotational transitions in the far-infrared. Even accounting for isotope-selective chemistry, \textit{HD can be assumed to have a fixed abundance relative to H$_{2}$ throughout almost the entire disc} \citep{trapman+2017}. The HD/H$_{2}$ ratio is determined by the local Galactic D/H ratio, which is $(2.0\pm0.1)\times10^{-5}$ \citep{prodanovic+2010}. Detecting the $J=1$--$0$ line with the \emph{Herschel} Space Observatory allowed \cite{bergin+2013} to estimate a total mass of 0.05~M$_\odot$ in the disc around TW~Hya (0.8~M$_\odot$ star). Later estimates, using more strongly constrained physical-chemical models and additional information from the HD $J=2$--$1$ transition, found ($0.075\pm0.015$)~M$_\odot$ \citep{trapman+2017}. HD detections have also yielded mass estimates for two other T~Tauri discs: DM~Tau (0.65~M$_\odot$ star) with $(2.9\pm1.9)\times10^{-2}$~M$_\odot$, and GM~Aur (1.1~M$_\odot$ star) with $(2.5\pm20.4)\times10^{-2}$~M$_\odot$ \citep{mcclure+2016}. The three measurements, with gas masses of 20-80~M$_{J}$, are consistent with a dust-to-gas mass ratio of 1:100 \citep{bohlin+1978}. 

It is important to note that particularly massive and bright discs were specifically targeted in these observations. \cite{kama+2020} have recently analysed the upper limits on the HD $J=1-0$ line flux of discs around intermediate mass stars putting a strong constraint on the gas mass of the disc around HD 163296, $\rm M_{disc}\leq 0.067\,$M$_{\odot}$. Comparing this with the masses of five candidate protoplanets in this disc, they find a giant planet formation mass efficiency of $\gtrsim 10\,$\% for present-day values. \textcolor{black}{Because of the moderate energy of the low-level  rotational lines (HD J=1, E$/k_B$ = 128.5\,K), the HD-based mass estimates rely on the knowledge of the disc thermal structure. This can be estimated comparing multiple transitions of optically thick lines with thermo-chemical models. An ideal disc "thermometer" is the CO rotational ladder \citep[e.g.][]{fedele+2016}. By fitting simultaneously the fluxes of the low-$J$ HD and multiple CO transitions with thermo-chemical models it is possible to obtain robust constraints to both the temperature structure and total gas mass \citep{trapman+2017}.}

\textit{CO is the second most abundant molecule after H$_2$}, with an abundance of $\le$10$^{-4}$ with respect to H$_2$, and its rotational emission lines are readily detected in the millimetre. Historically, many of the gas mass measurements have relied on such CO observations in the past, for the lack of ability to detect more reliable tracers with pre-ALMA instruments. Such \textit{measurements yield lower limits to the total gas mass reservoir}, as these bright \textit{emission lines are} also \textit{optically thick} and trace higher layers, and not the disc midplane where the bulk of the mass resides and planets form \citep{dartois+2003}. Another issue affecting CO is depletion due to freeze-out, as deep in the cold midplane CO readily freezes onto dust grains as soon as the temperature is below 20~K. \\
\indent \textit{Emission from CO isotopologues is less optically thick}, and in fact C$^{18}$O and C$^{17}$O are largely optically thin, making gas in the midplane accessible through millimetre observations. This method already yielded gas measurements with pre-ALMA instruments \citep{panic+2008} and is currently widely used. An important limitation is still the CO freeze out, and other depletion mechanisms such as selective photodissociation, due to which the CO abundance may be decreased below the commonly assumed values, thereby rendering such mass estimates lower limits only \citep[e.g.][]{dutrey+1996,ansdell+2016,miotello+2016}. \textcolor{black}{Depending on the disc surface density, even the commonly used C$^{18}$O lines can be optically thick in inner $\sim 10-20\,$au of the disc and the use of even rarer isotopologues is required. Recently, ALMA detected the rarest CO isotopologues  $^{13}$C$^{18}$O \citep{zhang+2017} and $^{13}$C$^{17}$O \citep{booth+2019,booth+2020}.}

\begin{tcolorbox}
It was long unclear whether unexpectedly low CO-based total disc mass estimates were due to an overall lower gas mass or a lack of CO molecules (``CO depletion''). Detailed physical-chemical models constrained by the continuum spectral energy distribution, multiple CO transitions and spatially resolved line maps, HD fluxes or upper limits, and other data have confirmed that \textbf{elemental C and O can have a wide range of gas-phase abundances, from nominal to two orders of magnitude depleted}, depending on the disc \citep{bruderer+2012,favre+2013,du+2015,kama+2016b,trapman+2017}. 
\end{tcolorbox}
\textcolor{black}{It should be noted that a measurement of the total optically thin CO isotopologue line flux by itself only yields a lower limit on the disc gas mass. Multi-line and continuum data can provide some insight into whether it is the CO abundance or the total gas mass which is low  \citep[e,g,][]{favre+2013,kama+2016b,woitke+2016,du+2017}. Recently, the rarest stable CO isotopologue $^{13}$C$^{17}$O has been detected in the disc HD~163296, yielding a gas mass of 0.3$M_{\odot}$, a few times higher than obtained with more abundant isotopologues \citep{booth+2019}. It is not yet clear whether the lower HD-based mass ($\leq0.067\,$M$_{\odot}$, \citealt{kama+2020}) is due to an enhancement of volatile abundances or something else.}

At the time of their formation, {\em discs inherit the dust-to-gas ratio from the molecular clouds}, and this fiducial value is often assumed to be 1:100 as measured in the ISM \citep{bohlin+1978}. \textit{Measuring the dust mass is reliable to roughly an order of magnitude}, which is much better than the uncertainties linked to the gas mass estimates, except in the rare cases when HD emission is available as a constraint. The mass locked in dust grains of order a millimetre to centimetre in size is calculated directly from the observed flux at a wavelength similar to the grain size, assuming optically thin emission and a cold temperature ($\approx20\,$K). For example, disc dust masses extend from $10^{-5}$ to $<10^{-3}\,$M$_\odot$ in Lupus \citep{ansdell+2016}. The Lupus discs detected in CO 
lines typically have gas masses below $<10^{-3}\,$M$_\odot$ (i.e., less than a Jovian mass of gas) which implies dust-to-gas ratios over 1:100. 

\textcolor{black}{As it has been recently shown, even at millimeter wavelength, the dust emission might not be optically thin throught the whole disc extent and self-scattering might bias the dust mass estimates that should be therefore considered as a lower limit \citep[e.g.][]{zhu+2019}.} As discussed above, the few available HD-based disc gas masses are in the $10$ to $100\,$M$_{\rm Jup}$ range. The field is currently trying to establish whether the low CO-based masses are real or a consequence of depletion of gas-phase elemental C and O \citep{krijt+2018,schwarz+2018}. \textcolor{black}{Future observations at even longer wavelengths with e.g., the ngVLA, will help to better constrain the total dust mass \citep[e.g.,][]{Isella+2015}.} We expect substantial progress on understanding disc gas mass evolution and planet formation efficiency over the coming years, particularly by the time Ariel is due to fly. 

Average dust masses derived from millimetre emission fail to reach 10~{M$_\oplus$}. This said, the dust mass \textcolor{black}{just refers to the solid material presently in form of dust in the protoplanetary discs and} \textit{does not provide any indication of how much solid material is contained in form of rocks and larger bodies}. These contribute to a negligible fraction of the millimetre flux, dominated by the grains of comparable size to the wavelength. Dust mass measurements are improved by complementary measurements of the dust spectral index, which provides a better grasp of dust opacity - one of the key sources of uncertainty.

\subsection{Chemical composition and molecular inventory of protoplanetary discs}

\begin{figure}[t]
\centering
\includegraphics[width=\textwidth]{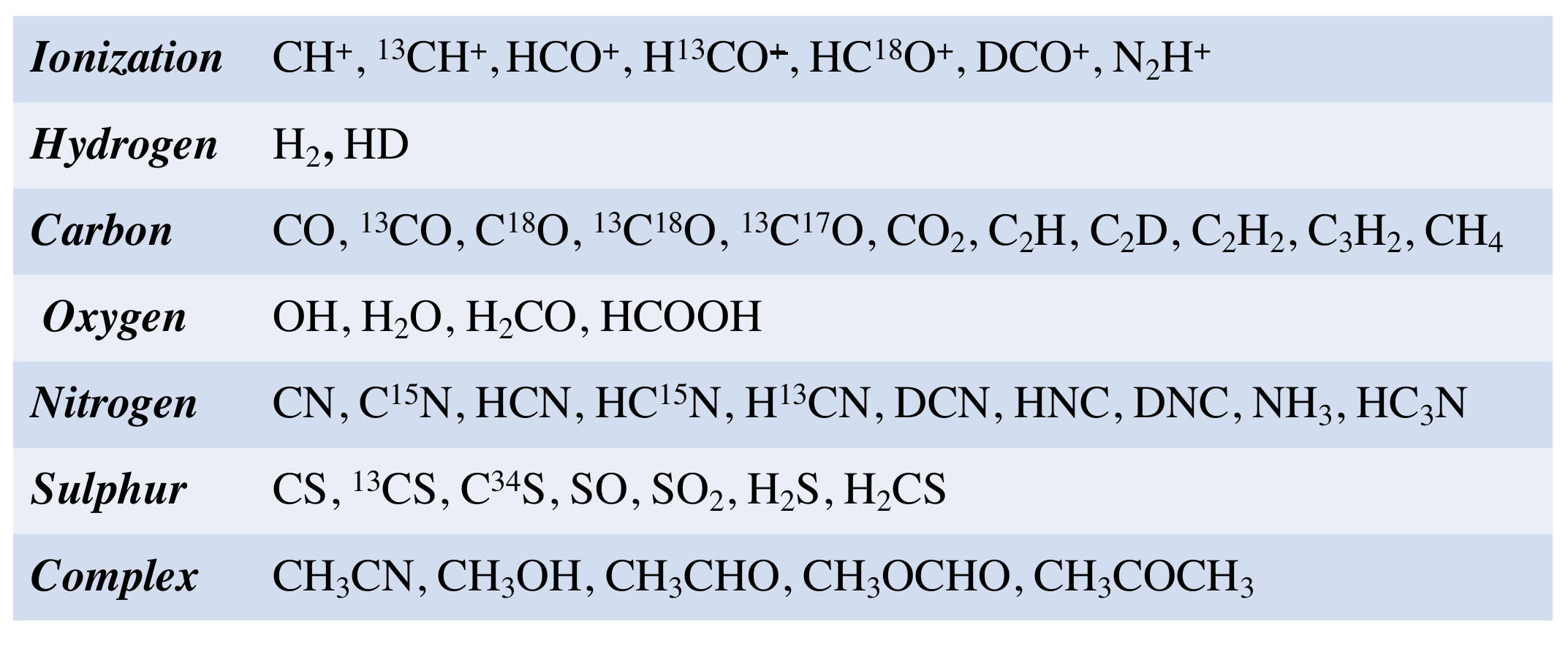}
\caption{List of molecules detected in protoplanetary discs at any wavelength. An increasing number of molecular isotopologues is being discovered at millimeter wavelenghts by ALMA. The detection of acetaldehyde (CH$_3$CHO), methyl formate (CH§$_3$OCHO) and acetone (CH$_3$COCH$_3$) refers to the outbursting source V883 Ori \citep{vanthoff+2018,Lee_JE+2019}.. The detection of C$_3$H$_2$ actually refers to c-C$_3$H$_2$, the cyclic form of this isomer whose linear form is yet to be detected in discs.}
\label{fig:inventory}
\end{figure}

\textcolor{black}{Figure~\ref{fig:inventory} summarises the inventory of molecules detected in discs (in the gas phase) at various wavelengths. Several carbon-, oxygen-, nitrogen- and sulphur-bearing species have been detected. 
Infrared observations of class II discs from the ground (with e.g., Keck/Nirspec and VLT/CRIRES) and from space (NASA/Spitzer and ESA/Herschel) revealed a molecule-rich inner disc. The most commonly detected species are OH and H$_2$O (e.g., \citealt{carr+2008, Mandell+2008, Pontoppidan+2010, Salyk+2011, Fedele+2011, Mandell+2012, fedele+2012, fedele+2013}). CO rovibrational emission has been a particularly powerful tracer of the structure of the inner disc and geometry of disc cavities and gaps (e.g., \citealt{Pontoppidan+2008, Brittain+2009, vanderplas+2009,Salyk+2011, Brown+2013}). A few simple organics molecules are also detected:  HCN, C$_2$H$_2$ and CO$_2$ (e.g., \citealt{carr+2008,Pascucci+2009,Pontoppidan+2010,Salyk+2011,Mandell+2012}) as well as CH$_4$ (seen in absorption in the disc around GV Tau N, \citealt{Gibb+2013}). These hot transitions emit from the inner ($\lesssim 10\,$au) warm molecular layer where molecules form primarily through gas-phase reactions (see e.g., \citealt{Woitke+2009, walsh+2015, Agundez+2018}). A further contribution can come from the thermal desorption of the ices within the H$_2$O snowline. The large column density of OH and H$_2$O requires an oxygen-rich inner disc. 
On the contrary, at longer wavelength, deep Herschel/HIFI observations of the ground state rotational transitions of H$_2$O resulted in only 2 detections in the discs around TW Hya and HD 100546, while the lines remain largely undetected in other discs \citep[][]{Hogerheijde+2011, du+2017} implying low abundance of cold H$_2$O ($\lesssim 10^{-11}$). The ground state H$_2$O emission is expected to come from the outer disc region ($> 50\,$au) where H$_2$O ice is photo-desorbed \citep[e.g.,][]{Woitke+2009}.
The emerging scenario is that, in the outer disc, oxygen is depleted onto icy grains, which release oxygen back to the gas-phase in the inner disc as they cross the water snowline, yielding an oxygen-rich inner disc \citep[see also][]{vandishoeck+2021} }

\smallskip
\noindent
Overall, the disc molecular inventory is made of a dozen of simple species. Complex molecules (i.e. composed of 6 or more atoms) are rare and the most complex species detected so far are: methanol (CH$_3$OH, \citealt{walsh+2016}), methyl cyanide (CH$_3$CN, \citealt{oberg+2015}) and formic acid (HCOOH, \citealt{favre+2018}). Recent observations of the FU Orionis-like object V883 Ori revealed the presence of more complex species such as CH$_3$CHO and CH$_3$COCH$_3$ along with CH$_3$OH \citep{vanthoff+2018,Lee_JE+2019}, opening the path to the investigation of complex species in discs. For most of the species, we do not have direct information of their radial and vertical distribution. Thanks to ALMA however, we are starting to spatially resolve the molecular emission, which informs us about the radial distribution of different volatiles.

\textbf{Outer disc.} On $10$ to $100\,$au scales, spatially unresolved gas-phase total elemental abundances or abundance ratios are, thus far, available for carbon (C), oxygen (O), nitrogen (N), and sulfur (S). These abundances are typically constrained by models including the disc physical structure, a chemical network, and ray-tracing of continuum and line emission. Sometimes, the vertically-integrated column density is the modelled quantity. In most cases, observational constraints include rotational lines of multiple species. The outer disc gas-phase elemental abundance and ratio measurements published to-date are summarized in Table~\ref{tab:elemratios}.

\textbf{Inner disc.} On $0.1$ to $10\,$au scales, in what was traditionally expected to be the formation zone of most planets before ALMA surveys revealed the possible signatures of giant planets at tens or even hundreds of au from the host stars \citep[e.g.][]{andrews+2018}, elemental and chemical abundances are very hard to determine due to the small emitting area and the poorly known physical structure of these regions. New techniques are focusing on studying the composition of the material accreting onto the central star as a proxy of the inner disc:
\begin{enumerate}
\item{Actively accretion-dominated photospheres of young stars above $1.4$ M$_{\odot}$  \citep{jermyn+2018}. Disc accretion rates are sufficient to cover the entire photosphere on weekly timescales in these stars, whose radiative envelopes are mixed with the deeper layers very slowly. Photospheric abundances provide a measurement of the absolute dust depletion level in gaps or cavities in the inner disc \citep{kama+2015}. This technique will in the near future allow to study variations in the mutual ratios of many elements, including C, N, O, S, Si, Fe, Mg, Al, Ca, and Ti. A recent result of its application has been the determination that $(89\pm8)\,$\% of all sulfur atoms in the inner few au of discs are locked in a refractory component, likely FeS \citep{kama+2019}\textcolor{black}{, confirming the picture depicted by the data provided by the meteorites and minor bodies in the inner \citep[e.g.][and references therein]{palme+2014,lodders+2019} and outer \citep[e.g.][and references therein]{altwegg+2019,rubin+2020} Solar System.}} 
\item{Atomic emission lines from the inner accretion disc where all dust has evaporated \citep{ardila+2013}. While these lines are hard to relate to absolute abundances, C and Si line emission can constrain major disc features such as the re-appearance of volatile carbon (reduced in outer disc gas by $\approx50\times$) within the dust-depleted inner cavity in TW Hya \citep{kama+2016b,mcclure+2019} or dust-evolution driven time variability in the inner disc volatile composition \citep{booth+2017}.}
\end{enumerate}

\begin{table}[t]
\centering
\caption{Gas-phase elemental abundances in discs on $10$ to $100\,$au scales.}
\begin{tabular}{ l c c c c c }
\hline\hline
Object & C/H & C/O & N/O & S/H & References \\
& ($\times10^{-4}$) & & & ($\times10^{-5}$) & \\
\hline
Sun & $2.69$ & $0.55$ & $0.16$ & $1.32$ &  \\
\hline
DM Tau & $0.2\ldots1.0$ & $>1$ &  & $<10^{-2}$ & 1,2,3  \\
GM Aur & $10^{-2}$ &  &  &  & 1 \\
GO Tau &  & $>1$ & & $<10^{-2}$ & 2 \\
HD 100546 & $1.35$ & $<0.9$ &  & $^{*}10^{-4}$ & 4 \\
 & $0.135$ & $<0.9$ &  & $^{*}10^{-3}$ & 5 \\
IM Lup &  & $0.8$ & $10$ &  & 6 \\
LkCa 15  &  & $>1$ &  & $<10^{-2}$ & 2 \\
TW Hya & $0.01$ & $>1.1$ &  & $^{*}10^{-4}$ &  5,7,8 \\
\hline
\end{tabular}
\newline
\small{\emph{Notes.} Reported abundances with respect to H nuclei may have systematic uncertainties of a factor of a few. The difference between C/O~$<1$ and $>1$ is strongly constrained and largely unrelated to these absolute scalings. \emph{References}: 1: \cite{mcclure+2016}; 2: \cite{dutrey+2011}; 3: \cite{semenov+2018}; 4: \cite{kama+2016a}; 5: \cite{kama+2016b}; 6: \cite{cleeves+2018}; 7: \cite{favre+2013}; 8: \cite{trapman+2017}; *: upcoming publications.}
\label{tab:elemratios}
\end{table}

\section{The Influence of the Stellar and Galactic Environments}\label{sec:galactic_environment}

Planetary systems do not form and exist in isolation, but instead are subject to radiative and dynamical influences from their cosmic environment. The birth of stars and planets takes place at local density peaks in the hierarchically structured \textcolor{black}{ISM} (e.g.~\citealt{krumholz+2005,kennicutt+2012}). The fractal nature of the ISM causes star formation to be spatially clustered \citep[e.g.][]{elmegreen+1996a,kruijssen+2012,hopkins+2013}. In other words, planetary systems are often born and sometimes evolve in the vicinity of other stars and planetary systems, which can have an important impact on their properties (see \citealt{adams+2010} and \citealt{kruijssen+2019} for recent reviews). 

The two dominant, external physical mechanisms that are generally considered to affect the properties of planetary systems are:
\begin{enumerate}
    \item {\it external photoevaporation} by massive stars, which accelerates the dispersal of protoplanetary discs and potentially modifies their chemistry and thermal structure, including the location of the snowlines  \citep[e.g.][]{scally+2001,winter+2018a};
    \item {\it dynamical encounters} with other stars, which can perturb either the protoplanetary disc or, on longer timescales, disrupt the planetary system itself \citep[e.g.][]{marzari+2013,davies+2014,rosotti+2014,bhandare+2016}.
\end{enumerate}
Observational indications of these mechanisms have been found. For instance, observations show a variation of the protoplanetary disc size with ambient stellar density \citep{dejuanovelar+2012}, a variation of the protoplanetary disc mass with distance to the nearest massive star \citep{ansdell+2017}, a decline of the fraction of stars with discs in clusters at increasing values of local UV fluxes \citep{guarcello+2016}, and tidal features as evidence of past dynamical encounters between protoplanetary discs \citep{winter+2018b}. It requires little imagination to realise that these effects can transform the architecture of the resulting planetary systems, as well as affect the bulk composition of the planets and that of their atmospheres.

It depends on the observable quantity of interest and on the timescale considered whether external photoevaporation or dynamical encounters dominate. When only concerned with protoplanetary disc dispersal, external photoevaporation is almost always the dominant external mechanism, except in (rare) cases of high stellar densities and no massive stars \citep{winter+2018a,winter+2020}. However, in the context of Ariel, we are not only interested in the truncation of the planet formation process, but also in the effect of external photoevaporation and irradiation on planet formation through changes in the chemistry and/or the thermal environment of discs. These processes require much less extreme circumstances (e.g.\ radiation fields of a few $100~{\rm G}_0$, \citealt{ciesla+2012}) than externally accelerated disc dispersal \citep[$>10^3~{\rm G}_0$,][]{winter+2020}. For instance, the synthesis of complex organic molecules and amino acids on icy dust grains is expected to be accelerated under external UV or soft X-ray irradiation \citep[e.g.][]{munozcaro+2002,throop+2011,ciaravella+2019}. Likewise, the formation of planetesimals may be accelerated by an external UV field \citep[e.g.][]{shuping+2003,throop+2005}. For both of these examples, it is plausible that the effect of external UV irradiation is a runaway process, because small dust grains can get trapped in photoevaporative flows, causing a decrease of the extinction column and a corresponding loss of UV absorption \citep[e.g.][]{balog+2006}.

When concerned with the architecture of planetary systems over long ($\sim$Gyr) timescales, the expectation value for the number of disruption events by dynamical encounters increases linearly with the age of the system, under the assumption that the ambient environment does not evolve. Therefore, eventually the integrated impact of dynamical encounters is expected to outweigh that of external photoevaporation and to become the dominant form of external perturbation for older planetary systems. In this context, it is critical to consider the gravitational boundedness of the birth stellar population, because only gravitationally bound clusters can generate sustained dynamical perturbations -- unbound stellar associations never live beyond a dynamical crossing time \citep{gieles+2011}, implying that encounters are an extreme rarity. In the current solar neighbourhood, about 5--10$\%$ of stars are born in bound clusters \citep[e.g.][]{lada+2003}, but this is expected to have been much larger in the past, with up to 50$\%$ of all stars and planetary systems with ages $>8$~Gyr having been born in bound clusters and potentially having been exposed to extreme external disruption \citep[e.g.][]{kruijssen+2012,longmore+2014,pfeffer+2018}. Notwithstanding these arguments, a sufficiently dense field star population can rival that of stellar clusters (e.g.\ towards galactic centres), such that even the field can generate a significant number of dynamical encounters over a sufficiently long timescale.

\begin{tcolorbox}
The above discussion has a number of crucial implications for the Ariel target selection and, more generally, for studying the link between planetary composition and formation environment. Above all, the fact that \textbf{the stellar and galactic environment beyond the confines of a planetary system can affect its architecture, composition, chemistry, and atmospheric properties}, implies that the target selection should aim for a diversity of possible formation environments.
\end{tcolorbox} 

\textcolor{black}{Major efforts are currently being undertaken to link the properties of planetary systems to their large-scale stellar environment, with promising results \citep[e.g.][]{winter+2020b,kruijssen+2020,longmore+2021,chevance+2021}. These studies show that planetary system architectures and planetary properties (e.g.\ multiplicity, semi-major axis distribution, Hot Jupiter incidence, planet radius distributions and uniformity) exhibit intriguing environmental dependences, which can be isolated most clearly for host stellar ages of 1-4.5~Gyr \citep{winter+2020b}. While it is not yet possible to unambiguously relate the current environmental conditions to those of the formation environment, there exist statistical trends that can already greatly inform the target selection of Ariel.}

\textcolor{black}{Specifically, the ambient stellar density of planetary systems at birth sets the strength of the external UV field, as well as the rates of external photoevaporation, dynamical encounters, and nearby supernovae, affecting isotopic abundance ratios \citep[e.g.][]{fujimoto+2018}. The birth density greatly increases with gas pressure and therefore age (due to cosmic evolution, up to about $\sim8$~Gyr ago, \citealt{kruijssen+2012,adamo+2015,pfeffer+2018,winter+2020}), which is likely to result in age trends of chemistry and atmospheric properties. Additionally, we may expect a trend with host stellar mass, because the impact of external radiative effects increases towards lower host stellar masses, due to lower binding energies and gas pressures of the protoplanetary discs \citep[e.g.][]{haworth+2018,winter+2020}. These trends are accessible by Ariel within the solar vicinity, due to the wide range of ages (and thus cosmic formation environments, see e.g.\ \citealt{ruizlara+2020}) and host stellar masses of the local population of planetary systems.}

The above line of reasoning leads to the following recommendation. In addition to the well-documented internal, secular processes governing the formation and evolution of planetary systems, 
the cosmic formation environment is a major axis along which Ariel's planet sample should be expected to reveal new, surprising, and physically important trends. 
To realise this discovery potential, Ariel should aim to target a sufficiently large sample of planets around low-mass stars, with a wide range of ages from \textcolor{black}{comparatively young} ($\sim1$~Gyr) to older ($>5$~Gyr), because \textcolor{black}{older} planetary systems around low-mass stars are predicted to be most strongly influenced by environmental effects. In addition to this general guideline, it is to be expected that the ongoing efforts aimed at characterising the formation environments of at least some of the known exoplanetary systems will bear fruit before the Ariel target selection has been finalised.

\section{The Importance of Stellar Characterisation}\label{sec:stellar_characterizarion}

Today we already know that planetary systems form around different types of stars, including low-mass stars like our Sun and more massive stars. Planets were found around both main sequence stars and evolved stars, and even around compact objects left from supernova explosions, like pulsars \citep{wolszczan1992}.
Thanks to Ariel we will gain new fundamental data on planets as they are today, but crucial properties related to their formation environments in the discs are long gone and cannot be observed anymore. On the other hand, part of this fundamental information is still available and preserved in the host stars. 
\subsection{Stellar host mass, type and metallicity: State of the Art from available observations}\label{sec: stars}

An important stellar property is its chemical composition. The iron abundance, expressed as [Fe/H] (the $\log$ number abundance of Fe/H relative to solar), is frequently used as a proxy for the metal content of the star. Already from the early studies of just four systems \cite{gonzalez+1997} noted that giant planets tend to orbit around metal-rich stars. It is well established that the frequency of gas-giant planets
(whose planetary mass M$_{\rm p}$ $>$ 30 M$_{\oplus}$) correlates with the stellar metallicity \citep{santos+2001,fischer+2005,sousa+2008,johnson+2010}. 
While the Sun and other nearby solar-type stars have [Fe/H] $\sim$ 0, typical exoplanet host stars
have  [Fe/H] $>$ 0.15.  Data shows that the percentage of stars with detected Jupiter-like planets with orbital periods less than 4 yr rises with the iron abundance from less than 3\% for the FGK stars with subsolar metallicity, up to 25\% for stars with [Fe/H] $\ge$ +0.3 dex \citep{fischer+2005}. 
On average, the metallicity distribution of stars with giant planets is shifted by 0.24 dex relative to that of stars without planets.

In contrast to the giant Jupiter-mass planets, less massive planets (either more similar to Neptune or super-Earths) do not form preferentially in higher metallicity environments  \citep{udry+2006,sousa+2008,ghezzi+2010a,mayor+2011}. Indeed, the median metallicity for solar-type stars hosting low-mass planets is close to -0.10 dex, and a significant number of low-mass planets are orbiting around stars with metallicities as low as -0.40 dex \citep{mayor+2011}. This observational result is usually explained within the framework of core-accretions models \citep[e.g.][]{pollack+1996,rice+2003,alibert+2004}
which assume that the timescale needed to form an icy/rocky core is largely dependent on the metal content of the protostellar cloud. In this way, in low-metal environments, the gas has already been depleted from the disc by the time the cores are massive enough to start a runaway accretion of gas. As a result, only low-mass planets can be formed. A possible correlation between planetary mass and stellar metallicity have been also suggested by \cite{jenkins+2017}. 

The correlation between planetary occurrence and metallicity observed in main sequence stars may not extend to giant stars, as several studies have found contradictory results 
\citep[e.g.][]{pasquini+2007,ghezzi+2010a,maldonado+2013,mortier+2013,jofre+2015,reffert+2015}. Several explanations have been put forward to explain the possible lack of a planet-metallicity correlation in evolved stars including the accretion of metal-rich material, higher-mass protoplanetary discs, and the formation of massive gas-giant planets by metal-independent mechanisms. 
\textcolor{black}{Note that red giants are the result of the evolution of MS dwarfs with spectral types G5V-B8V (stellar masses between 0.9 and 4 M$_{\odot}$).
High-mass stars are likely to harbour more massive protoplanetary discs 
\citep[][]{alibert+2011,muzerolle+2003,natta+2004,mendigutia+2011,mendigutia+2012}.
Simulations of planet population synthesis \citep[][]{alibert+2011,mordasini+2012}
show that giant planet formation can occur in low-metallicity (low dust-to-gas ratio) but high-mass protoplanetary discs. This effect depends on the mass of the disc. The minimum metallicity required to form a massive planet is lower for massive stars than for low-mass stars. In this scenario, the fact that giant stars with planets do not show the metal-rich signature could be explained by the more massive protoplanetary discs of their progenitors. }
Planets around evolved stars show some peculiarities with respect to the planets orbiting around main-sequence stars, like a lack of close-in planets or higher masses and eccentricities \cite[e.g.][]{maldonado+2013}. There is also a strong dependence of giant planet occurrence on stellar mass: stars of $\sim$1.9 M$_{\odot}$ have the highest probability of hosting a giant planet \cite[][]{reffert+2015}. 

More observational evidence has been reported presenting correlations between planetary radius and host star metallicity \cite[][]{buchhave+2014, schlaufman+2015}, as well as correlations between the eccentricity of planets versus metallicity \cite[][]{adibekyan+2013, wilson+2018}. Although no clear correlations have been found between the stellar metallicity and the planetary semi-major axis, recent works discussed whether the stellar hosts of hot Jupiters (a $<$ 0.1 au) show higher metallicities than stars with more distant planets. For example, \cite{maldonado+2018} shows that the metallicity distribution of stars with hot gas-giant planets is shifted by 0.08 dex with respect to that of stars with cool distant giants. The authors also noted a paucity of hot Jupiters orbiting stars with metallicities below -0.10 dex, whilst cool Jupiters can be found around more metal-poor stars. Along these lines, \cite{bashi+2017} and \cite{santos+2017} 
suggest that stars hosting massive gas-giant planets show on average lower metallicities than the stars hosting planets with masses below 4–5 M$_{\rm Jup}$. Finally, unlike gas-giant planet hosts, stars with brown dwarfs do not show metal enrichment \cite[e.g.][]{ma+2014} although \cite{maldonado+2017} found that stars with low-mass brown dwarfs tend to show higher metallicities than stars hosting more massive brown dwarfs.

\begin{tcolorbox}
The emerging picture is one in which \textbf{different planet formation mechanisms may operate altogether and their relative efficiency changes with the mass of the substellar object that is formed}. For substellar objects with masses in the range from 30 M$_{\oplus}$ up to several Jupiter masses, high host star metallicities are found, suggesting that these planets are mainly formed by the core-accretion mechanism. More massive substellar objects do not tend to orbit preferentially around metal-rich stars and are likely to be formed by gravitational instability or gravoturbulent fragmentation \citep[][]{schlaufman+2018,maldonado+2019}.
\end{tcolorbox}

\textcolor{black}{Most of the planet-host stars with low iron content  belong to the thick disc population \citep{haywood+2008,adibekyan+2012}. Thin-disc stars rotate faster than the local standard of rest and show solar $\alpha$-element abundances (Mg, Si, Ca, Ti). On the other hand, the thick disc is enriched in alpha elements and lags behind the local standard of rest \citep{reddy+2003,reddy+2006}.} It has been argued that to form a sufficiently massive core, the quantity that should be considered is the surface density of all condensible elements beyond the ice line \citep{mordasini+2012}, especially the elements O, Si, and Mg \citep{robinson+2006,gonzalez+2009}. In particular, Mg, and Si have condensation temperatures very similar to Fe \citep{lodders+2003}.  It is therefore likely that stars with intermediate metallicity might compensate their lower metal content with other contributors to allow for planet formation to occur. Along these lines, \cite{adibekyan+2012} found that most of the planet-host stars with low Fe content are enhanced by $\alpha$ elements. This $\alpha$-enhancement is a common property of most of metal-poor stars and is an effect of galactic chemical evolution, as most of the present Fe is made in thermonuclear supernovae while most of the $\alpha$-elements like O or Si are made mostly in massive stars.

Regarding other chemical abundances besides Fe, planet-hosting stars are largely indistinguishable in their enrichment histories of refractory elements \cite[e.g.][]{bodaghee+2003,gilli+2006,adibekyan+2012}, or show rather modest overabundances, with respect to other stars without planets. Volatile elements (C, O, Na, S, and Zn) are important in the chemistry of protoplanetary discs and planets. Stellar abundances can be difficult to estimate and high resolution data with high signal-to-noise are necessary to quantifying them. Spectral lines can be weak and blended, depending by the specific element, and cooler stars (T $<$ 4500 K) in particular present molecular bands that need a specific approach for treating elemental abundances. Interestingly, most volatiles show a decreasing trend of [X/Fe] with increasing [Fe/H], but the abundance trends for planet-hosting stars for volatile elements are similar to those for the comparison stars at the corresponding (high) values of [Fe/H] \cite[e.g.][]{ecuvillon+2004,ecuvillon+2006a,gonzalez+2006,dasilva+2011}. More references can be found in \textcolor{black}{\cite{Perryman2018}}. 

The abundance of lithium is an important diagnostic of stellar evolution. Several works have suggested that stars with planets tend to have less lithium than stars without. In particular,  \cite{israelian+2004} found an excess Li depletion in planet-hosting stars with effective temperatures in the range 5600-5850 K, but no significant differences at higher temperatures. While this result has been confirmed by the majority of studies \citep[e.g.][]{gonzalez+2008,gonzalez+2010a,delgadomena+2014,delgadomena+2015}, other works have reported an absence of depletion for planet-hosting stars \citep[e.g.][]{ghezzi+2010a,ramirez+2012}. High Li abundance has also been reported in several rapidly rotating red giants that might be attributed to recent planet engulfment \citep[e.g.][]{adamow+2012,adamow+2014,nowak+2013}.

Beryllium, like lithium, is another tracer of the internal structure and (pre) main-sequence evolution. A higher beryllium depletion has been found for stars with effective temperatures lower than 5500 K, but this process seems to be independent of the presence of planets \citep[][]{delgadomena+2012}. Detailed chemical abundances of planet hosts has shown that the Sun and other solar analogues are depleted on refractory relative to volatile elements by $\sim$0.08 dex when compared with the majority of nearby solar twins \citep[e.g.][]{melendez+2009,ramirez+2009,ramirez+2010,ramirez+2014,gonzalez+2010a,gonzalez+2011}. After discussing several possible origins, \cite{melendez+2009} conclude that the most likely explanation is related to the formation of planetary systems like our own, in particular to the formation of rocky low-mass planets. Although appealing, this hypothesis has been questioned, and other works point towards chemical evolution effects \citep[][]{gonzalezhernandez+2010,gonzalezhernandez+2013,schuler+2011,maldonado+2015}, or an inner Galactic origin of the planet hosts \citep[e.g.][]{adibekyan+2014,maldonado+2016} as their possible causes.

One aspect that is currently challenging the detection of low-mass planets is stellar activity \cite[e.g.][]{dumusque+2012}. Exoplanet host stars display a wide variety of chromospheric and magnetic activities dependent mostly on their spectral type \citep[e.g.][and references therein]{Perryman2018}. For example, \cite{isaacson+2010} and \cite{arriagada+2011} performed a detailed study of large samples of stars in planet search programs, using activity indexes to estimate the level of radial velocity jitter of the program stars. \cite{cantomartins+2011} compared the activity index R$^{\rm '}_{\rm HK}$ measured in stars with and without planets finding similar distributions. In addition, no significant correlations between R$^{\rm '}_{\rm HK}$ and the planetary properties were found. \cite{gonzalez+2011} found that stars with planets have significantly smaller values of v$\sin i$ and R$^{\rm '}_{\rm HK}$  compared to otherwise similar non-planet hosts. No differences in the X-ray emission between planet hosts and non-host were found by \cite{kashyap+2008} although the authors reported higher X-ray luminosities for the stars hosting close-in giant planets. However, \cite{poppenhaeger+2010} found no correlation between the X-ray luminosity and the planet's mass or orbital distance. 

Observations indicate that host stars of some close-in hot Jupiters undergo episodes of periodic or enhanced stellar activity, linked to the presence of the planet through magnetic or tidal interactions \citep[see e.g.][and references therein]{Perryman2018}. Periodic activity has been reported both in the Ca~{\sc ii} H \& K and/or Balmer line emission in several planet-hosting stars, such as $\nu$ And and HD 179949, and inferred from optical brightness variations in the case of $\tau$ Boo  \citep{scandariato+2013,shkolnik+2005,walker+2008}. Another example is HD 17156, where enhanced chromospheric and coronal emissions were detected a few hours after the passage of the planet at the periastron \cite[][]{maggio+2015}. \\

\begin{figure*}[t]
\centering
\includegraphics[trim= 0.3cm 0cm 1.3cm 1.3cm, clip, width=0.495\columnwidth]{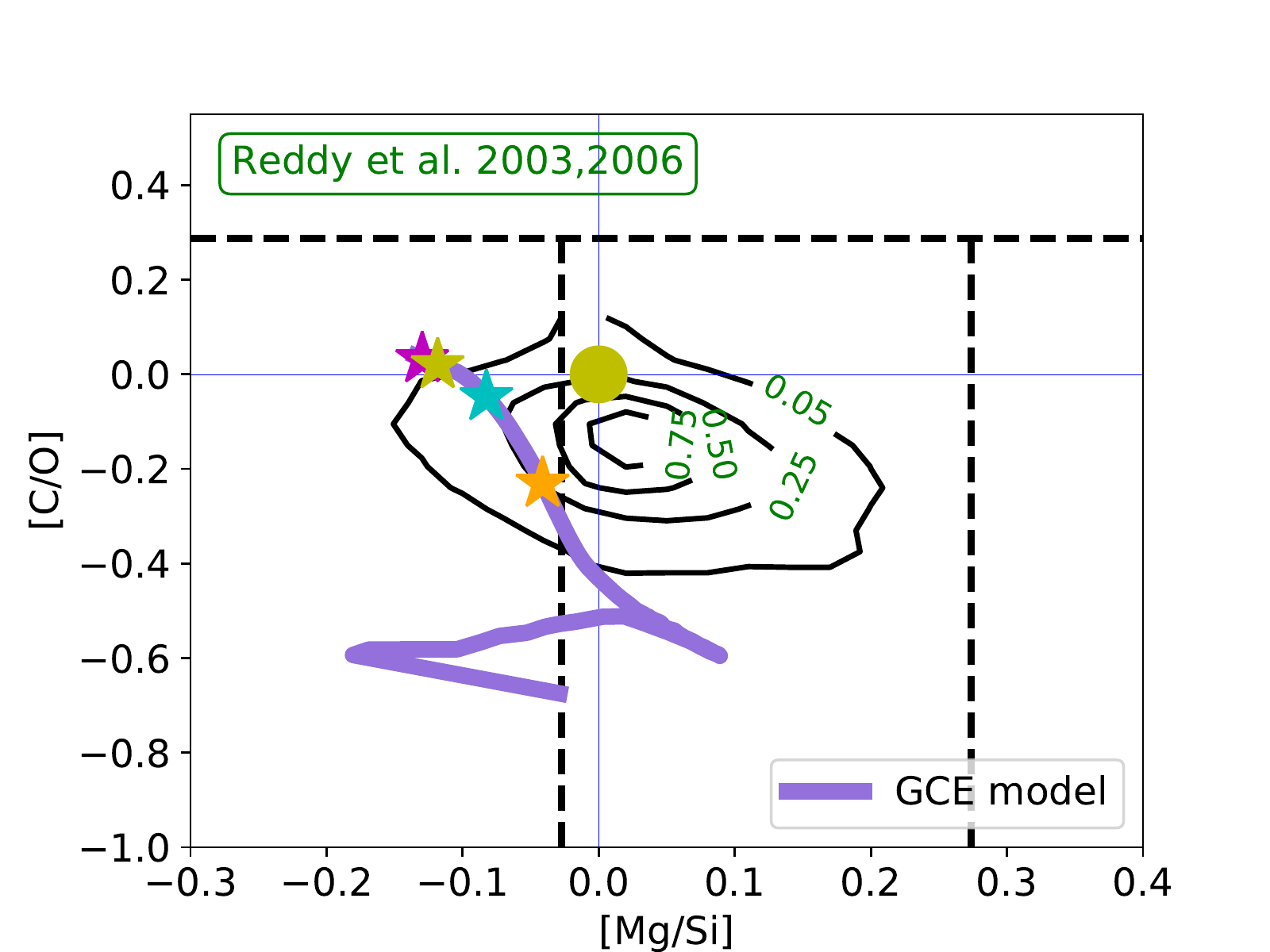}
\includegraphics[trim= 0.3cm 0cm 1.3cm 1.3cm, clip, width=0.495\columnwidth]{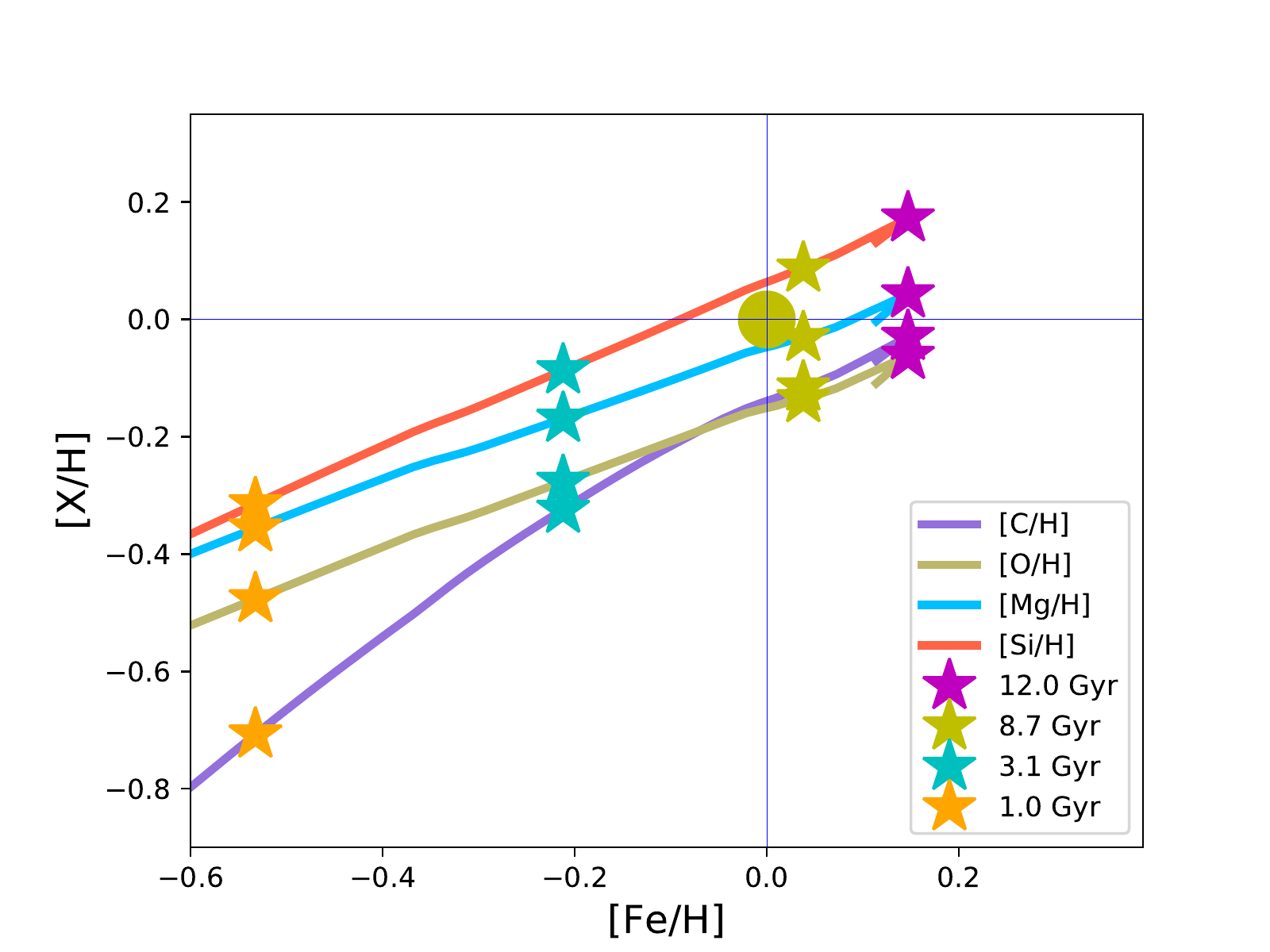}
\caption{\emph{Left Panel:} stellar statistics distribution with respect to [C/O] and [Mg/Si] from observations \cite{reddy+2003,reddy+2006} and GCE models. The elemental ratios are reported in logarithmic notation, scaled to solar value. The Sun is reported as reference, with blue lines indicating the solar ratios. Dashed black lines indicate the element classification relevant for planet behavior according to \cite{bond+2010}. GCE simulations are reported (magenta line), with reference time symbols at 1 Gyr, 3.1 Gyr, 8.2 Gyr (which corresponds to the age of the Sun) and 12 Gyr after the formation of the Milky Way (orange, cyan, yellow and magenta star, respectively). \emph{Right Panel:} [C/H], [O/H], [Mg/H] and [Si/H] are reported with respect to [Fe/H] for the same GCE model reported in the left panel.}
\label{fig: GCE_vs_disc}%
\end{figure*}

\subsection{Pristine chemical composition of the star and planet system}\label{sec: chem}

The original chemical composition of a stellar system, encompassing the star and its circumstellar disc, is part of the initial conditions for the planet formation process and sets some of the fundamental properties of the planets that the star will host. In particular, the initial chemical setup of the system will affect the envelope/mantle properties and the atmospheric composition of the planets. The relative ratios of elements such as O, C, Mg and Si will set the mineralogy of the planets and the impact of volcanic activity on their atmospheres \citep[e.g.,][]{bond+2010,delgadomena+2010}. 

The information on the initial chemical composition of the system will be modified and partly erased throughout the planet formation process (e.g. due to the different behaviour of volatile and refractory elements and the effects of planetary migration, see Sects. \ref{sec:giant_planets} and \ref{sec:high_density_planets}). The surface/photospheric composition of the host star, however, will preserve this information almost unchanged over time \cite[e.g.,][]{piersanti+2007}. The characterization of the stellar composition, therefore, provides fundamental information and the context to reconstruct the dynamical and formation histories of planets from their observed atmospheric composition (see also Sect. \ref{sec:giant_planets}). While the definition of the Solar System abundances (both present and protosolar) is constantly updated thanks to photospheric and meteoritic data, and today we know them with good precision \citep[though uncertainties still remain, see e.g. the oxygen crisis, ][and references therein]{palme+2014,lodders+2019}, the same is not true for all stars.

\begin{tcolorbox}
For most stars many elements are either not available or observed with much larger errors than Fe, and other elements like phosphorus and chlorine are not observable. Furthermore, the Milky Way disc is a complex system, where galactic chemical evolution (GCE) is changing chemical abundances over a timescale of billions of years, and the chemical enrichment history among stars with similar age and/or galactic location may show significant variations beyond observational errors. For this very reason, \textbf{the solar initial composition cannot be universally applied to different stellar systems}. 
\end{tcolorbox}

The characterization of the stellar host composition and of the initial planetary system composition needs to adopt a multi-disciplinary approach: available elemental spectroscopic data can be integrated with theoretical GCE simulations. In figure~\ref{fig: GCE_vs_disc}, left panel, the elemental ratios [C/O] and [Mg/Si] are shown for a stellar sample of the Milky Way disc by \cite{reddy+2003, reddy+2006}, covering a metallicity range similar to that of the Ariel expected targets \citep[][]{edwards+2019}. The Sun is also reported for comparison. Without considering 5$\%$ of disc stars with more exotic composition, the remaining 95$\%$ stars show a variation of [C/O] and [Mg/Si] between 0.3 and 0.4 dex in logarithmic notation, which corresponds to a variation between a factor of 2 and 2.5. GCE simulations for one possible chemical enrichment history are also shown in Fig. \ref{fig: GCE_vs_disc} for comparison.

The model was generated using the GCE code OMEGA from the NuPyCEE package \cite[][]{ritter+2016,cote+2017}. The yellow star symbol corresponds to the Sun formation time, 4.6 billions years ago. The model shown seems to provide an acceptable representation of the Sun, with a perfect match of the [C/O] ratio, and underestimating by about 0.1 dex (this corresponds to 25\%) the [Mg/Si] ratio. The chemical enrichment history from the same model is shown in Fig.~\ref{fig: GCE_vs_disc}, right panel, for C, O, Mg and Si, in the relevant metallicity range [Fe/H] currently foreseen for the Ariel sample \citep[][]{edwards+2019}. Within the metallicity range considered, it is interesting to observe the different trend of Si and C compared to O and Mg. This is due to the different chemical enrichment history of these elements \citep[e.g.,][]{timmes+1995a,kobayashi+2011,mishenina+2017}.

For a benchmark of the correct initial composition of all Ariel targets, an extended set of consistent GCE models will need to be generated covering the observations for element ratios like [C/O] and [Mg/Si], at the correct age of the stellar hosts. These models will have all the elements available, at any time, to compare with stellar observations and inform theoretical planet simulations. The relative abundance scatter obtained from these simulations need to be simulated, and their impact on planet simulations is still unknown.  

\subsection{Pristine radioactivity, local enrichment and galactic chemical evolution}\label{sec: radio}

Together with stable isotopes and elements, stars also make radioactive isotopes. Some of them have an half-life long enough to affect the formation and evolution of new-forming planetary systems. The $^{26}$Al and $^{60}$Fe isotopes have a half-life of 7.17$\times$10$^5$ years and 2.62$\times$10$^6$ years, respectively. Traces of extinct $^{26}$Al and $^{60}$Fe were found in the early Solar System condensates, and we know that their heat contribution is crucial during planetesimal formation \cite[][and references therein]{lugaro+2018}. The short half-life of $^{26}$Al and its abundance derived from early Solar System material exclude a relevant GCE contribution. Therefore, its abundance in the protosolar nebula is due to the contribution of local stellar sources, like the explosion of nearby supernovae, baking these species relatively nearby and shortly before the formation of the Sun. The argument is more controversial for $^{60}$Fe, where its low abundance relative to $^{26}$Al in the early Solar System is making unclear if its composition was a GCE product or if it was dominated from local stellar sources, as for $^{26}$Al \citep[][and references therein]{cote+2019a}.

Today we have evidence that pollution of radioactive material from other stars continued over time, not only during the formation of the Solar System. For instance, traces of the radioactive isotope $^{60}$Fe have been discovered in fresh Antarctic snow, signifying further pollution from recent supernovae \cite[][]{koll+2019}. Extinct $^{60}$Fe has been detected in the ocean crust, falling through Earth's atmosphere and oceans about 2.2 million years ago \cite[e.g.,][]{ludwig+2016}. This has been connected with a mass extinction event affecting many species in our oceans about 2.6 million years ago, due to the radiation from the same nearby supernova \cite[e.g.,][]{melott+2019}. This makes really hard to define what is the correct amount of $^{26}$Al and $^{60}$Fe to use in simulations of planet formation. 

Values obtained from meteorites for the early Solar System are the outcome of local stellar pollution or, perhaps, the combined outcome of stellar pollution and GCE for $^{60}$Fe. Even an ideal stellar system, composed by a solar twin with the exact same elemental composition, may have a complete different concentration of initial $^{26}$Al and $^{60}$Fe. And, as we mentioned before, GCE simulations alone are not providing a correct answer, since the half-life of these isotopes is too short. In preparation to Ariel's observations, a new generation of theoretical simulations will be needed. The initial $^{26}$Al and $^{60}$Fe abundances need to to be varied within a realistic range, from background GCE concentrations \cite[e.g.,][]{cote+2019a}, up to realistic abundances guided from stellar simulations of different types of stars that can produce $^{26}$Al and $^{60}$Fe \cite[e.g.,][]{timmes+1995b,limongi+2006,jones+2019,cote+2019b}.

The radioactive isotopes $^{40}$K, $^{232}$Th, $^{235}$U and $^{238}$U all have half-lifes of the order of 1 billion years or larger. Therefore, their galactic enrichment history behaves more like that of stable isotopes, and GCE simulations are needed to calculate their evolution in the interstellar medium. 
Once the planets are formed, the abundance of these species determines their internal heating history, sustaining tectonic activity and the magnetic field. The potassium observed today on the surface of the stellar host does not tell much about the initial $^{40}$K, and thorium and uranium are extremely hard to observe in stellar spectra. A possible strategy would be to get the realistic range of these abundances from GCE simulations, where the stellar sources of these isotopes are properly considered.

This analysis is more difficult for thorium and uranium, since they are product of the rapid neutron capture process (r-process). The dominant astrophysical source of the r-process is still matter of debate, and several sites like neutron-star mergers have been proposed \cite[e.g.,][and references therein]{cowan+2019,cote+2019c}.
An approach complementary to the one mentioned above is to perform a study of the production of $^{40}$K and actinide elements, to define observations from the abundance pattern measured in the stellar host that can be used as a diagnostic of their concentration. For instance, the r-process element europium measured from the surface of the stellar hosts could be used as a diagnostic for the initial abundance of the radioactive isotopes $^{232}$Th, $^{235}$U and $^{238}$U. At present, there is no available estimation of how much the abundances of these isotopes may change in the galactic disc in the metallicity range of interest for Ariel targets. 

\section{Organics as C-O-N Carriers}\label{sec:organics}

In the last 10 years there has been a number of surveys dedicated to assess the molecular content of protoplanetary discs, targeting mostly the simpler bi- and tri-atomic molecules such as CO, CN, H$_{2}$O, HCO$^+$, DCO$^+$, N$_{2}$H$^+$, HCN, DCN \citep[e.g.][]{fedele+2013,guilloteau+2013,guilloteau+2016,oberg+2010,oberg+2011,podio+2013}, \textcolor{black}{as well as a few S-bearing species, e.g. CS, H$_2$CS, H$_2$S \citep[e.g., ][]{phuong+2018,teague+2018,legal+2019,codella+2020,garufi+2020,garufi+2021,loomis+2020,vanthoff+2020}}. The content of  complex organic molecules (COMs), on the contrary, is still only poorly known because of their lower gas-phase abundances ($< 10^{-8}$ with respect to the abundance of atomic hydrogen). According to disc models this is due to the fact that COMs are frozen on the icy mantles of dust grains in the cold disc interior and only a tiny fraction of them is released in gas-phase through thermal or photo-/CR- desorption \cite[e.g.][]{aikawa+1999,willacy+2009,walsh+2014,loomis+2015}. Hence, (complex) organic molecules in protoplanetary discs remain hidden in their ices and can be unveiled only through interferometric observations at high sensitivity and resolution, e.g. with ALMA.
\begin{tcolorbox}
The study of COMs in protoplanetary discs is key to estimate the fraction of C, O, and N atoms that are trapped in organic molecules. While it is generally thought that most of these atoms are incorporated in ices (in the form of, e.g. H$_{2}$O, CO$_{2}$, CO, etc), there is the possibility that organic molecules trap a significant amount of them, as also suggested by recent observations of comets in the Solar System \cite[e.g.][]{fulle+2019}. \textbf{As organic molecules are carriers of C, O, and N, it is crucial to estimate their abundance and the location of  their snowlines, to infer accurate C/O/N ratios in the gas and solids of discs, with direct implications for the final C/O/N ratios of planets}.
\end{tcolorbox}

Thanks to ALMA a few simple organics have been imaged in discs. Among those, formaldehyde (H$_2$CO) and methanol (CH$_3$OH) are key to investigate organics formation. While H$_2$CO can form both in gas-phase and on grains, CH$_3$OH forms exclusively on grains. 
\textcolor{black}{An illustrative example are the resolved ALMA images of H$_2$CO in nearby protoplanetary discs \citep{qi+2013,loomis+2015,oberg+2017,carney+2017,carney+2019,podio+2019,garufi+2020,pegues+2020,podio+2020a,podio+2020b,garufi+2021}.} These allowed us to infer the H$_2$CO abundance and distribution in \textcolor{black}{protoplanetary} discs and to constrain the mechanism(s) for its formation \textcolor{black}{in this environment. This is key, given that H$_2$CO is} one of the bricks for the formation of complex organic and prebiotic molecules. 

\textcolor{black}{The distribution of H$_2$CO suggests that 
the bulk of the observed H$_2$CO in the disc is formed via gas-phase reactions. This is indicated both by the o/p ratios measured in TW Hya \citep{terwisschavanscheltinga+2021} and by the vertical distribution of molecular emission in the edge-on disc of IRAS 04302 where H$_2$CO mostly arises from an intermediate disc layer, the so called molecular layer \citep{podio+2020b,vanthoff+2020}.
However, for IRAS 04302 H$_2$CO emission is detected also in the outer disc midplane, where molecules are expected to be frozen onto the dust grain icy mantles. Also for a few discs the observations show a secondary emission peak of H$_2$CO located outside the CO snowline which may argue in favour of H$_2$CO formation on grains in these outer disc regions, following  
freeze-out of CO and its subsequent hydrogenation on the icy grains \citep{oberg+2017,carney+2017,podio+2019,pegues+2020}  (see  e.g. Fig. 3-Right). This may indicate} that ice chemistry is efficient in the outer regions of discs and \textcolor{black}{could} produce methanol as well as other complex organics, which are then partly released in gas-phase \textcolor{black}{via non-thermal processes. However,} only a few of them has been so far detected, i.e. cyanoacetylene and methyl cyanide (HC$_3$N, CH$_3$CN, \textcolor{black}{\citealt{oberg+2015,bergner+2018})}, methanol (CH$_3$OH, \textcolor{black}{\citealt{walsh+2016,podio+2020a}}), and formic acid (HCOOH, \citealt{favre+2018}). Typical abundances are: 10$^{-12}$-10$^{-10}$ (H$_2$CO), 10$^{-12}$-10$^{-11}$ (CH$_3$OH, HCOOH, HC$_3$N), 10$^{-13}$-10$^{-12}$ (CH$_3$CN).

\begin{figure}[t]
    \centering
    \includegraphics[width=\columnwidth]{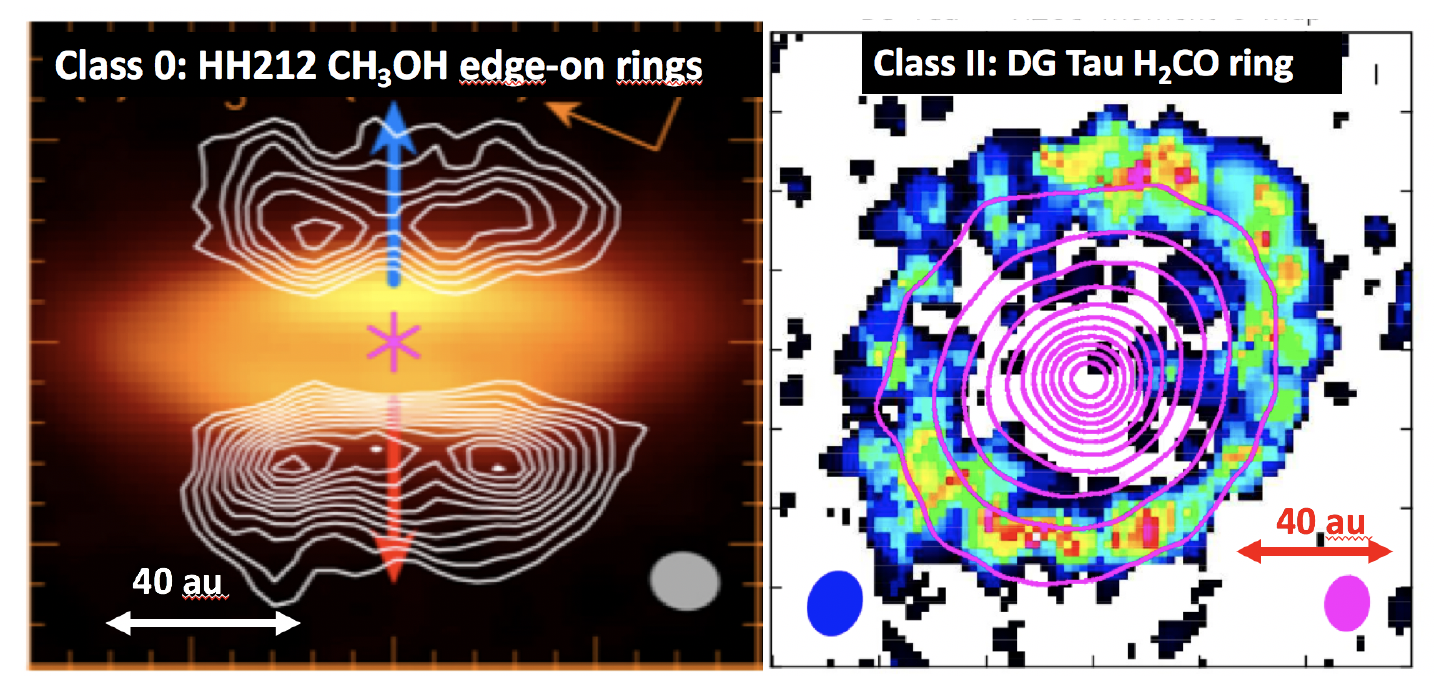}
    \caption{The protostellar discs around the Class 0 star HH 212 and the protoplanetary disc of the Class II star DG Tau as observed with ALMA. \emph{Left}: continuum (colour) and CH$_3$CO (contours) of the protostellar discs around HH 212. \emph{Right}: continuum (contours) and H$_2$CO (colour) of the protoplanetary disc of DG Tau. Figures adapted from \cite{lee+2019} and \cite{podio+2019}.}
    \label{fig:organics}
\end{figure}

At the protostellar stage observations are hindered by the presence of many kinematical components that may hide the chemical content of the disc, e.g. the surrounding envelope and the outflow. To date only one protostellar disc has been chemically characterised on Solar System scale, the disc of HH 212 (see Fig. 3-Left). This pioneering work shows enriched chemistry associated with the disc surface layers, with the detection of a number of complex organic molecules, e.g. CH$_3$OH (10$^{-7}$), HCOOH (10$^{-9}$), CH$_3$CHO (10$^{-9}$), HCOOCH$_3$ (10$^{-9}$), NH$_2$CHO (10$^{-10}$) \citep[e.g.][]{codella+2018,codella+2019,lee+2017,lee+2019}. These abundances are larger than what observed at the protoplanetary stage.

This chemical enrichment may be due to slow shocks occurring at the interface between the infalling envelope and the forming disc \cite[e.g.][]{sakai+2014}. The chemically enriched gas should then settle in the rotationally-supported disc, where the chemical composition is likely stratified and affected by the dynamics and the dust coagulation. The scenario obtained for HH 212 needs to be confirmed by collecting observations on a statistical sample. Answers are expected soon from the FAUST ALMA Large Program (Fifty AU STudy of the chemistry in the disc/envelope system of Solar-like protostars, \url{http://faust-alma.riken.jp}) and from the ALMA-DOT program (ALMA chemical survey of disc-outflow sources in Taurus, \citealt{garufi+2020}). The goal of these projects is to reveal the chemical composition of the envelope/disc system on Solar System scale in a sample of discs from the protostellar to the protoplanetary stages.

In order to test the inheritance scenario, i.e. whether the molecular setup of protoplanetary discs is largely inherited from the molecular clouds from which the stars formed, it is also crucial to compare the chemical composition of protostellar objects with that of Solar System objects (the final stage of Sun-like stars formation process). Comets are ideal for this purpose, as they sample the pristine composition of the outer Solar System. A comparative study of the comet 67P Churyumov-Gerasimenko, visited and characterized in detail by the ESA mission Rosetta, with two Solar-like protostellar systems, IRAS16293-2422B and SVS13-A, shows similar abundances of NH$_2$CHO and HCOOCH$_3$, and in general of CHO-, N- and S-bearing species, which suggests inheritance from the presolar phase \citep{bianchi+2019,drozdovskaya+2019}. Also in this case these promising results need to be confirmed by further comparative studies.

\section{Giant Planets and their Composition}\label{sec:giant_planets}

Giant planets, from Neptune-like to super-Jovian planets, currently represent the bulk of Ariel's observational sample \citep{edwards+2019}: as a result, efforts are being devoted to improve our understanding of what factors can shape their atmospheric composition as will be observed by Ariel \citep{turrini+2018}. In the following we focus on three of these factors: the link between orbital migration and metallicity (Sect. \ref{sec:metallicity}), the distribution of accreted material between the core and the growing envelope (Sect. \ref{sec:envelope}), and the compositional features arising from different migration histories (Sect. \ref{sec:tracers}).

\subsection{Planetary migration and bulk metallicity}\label{sec:metallicity}

\begin{figure}
\centering
\includegraphics[width=\columnwidth]{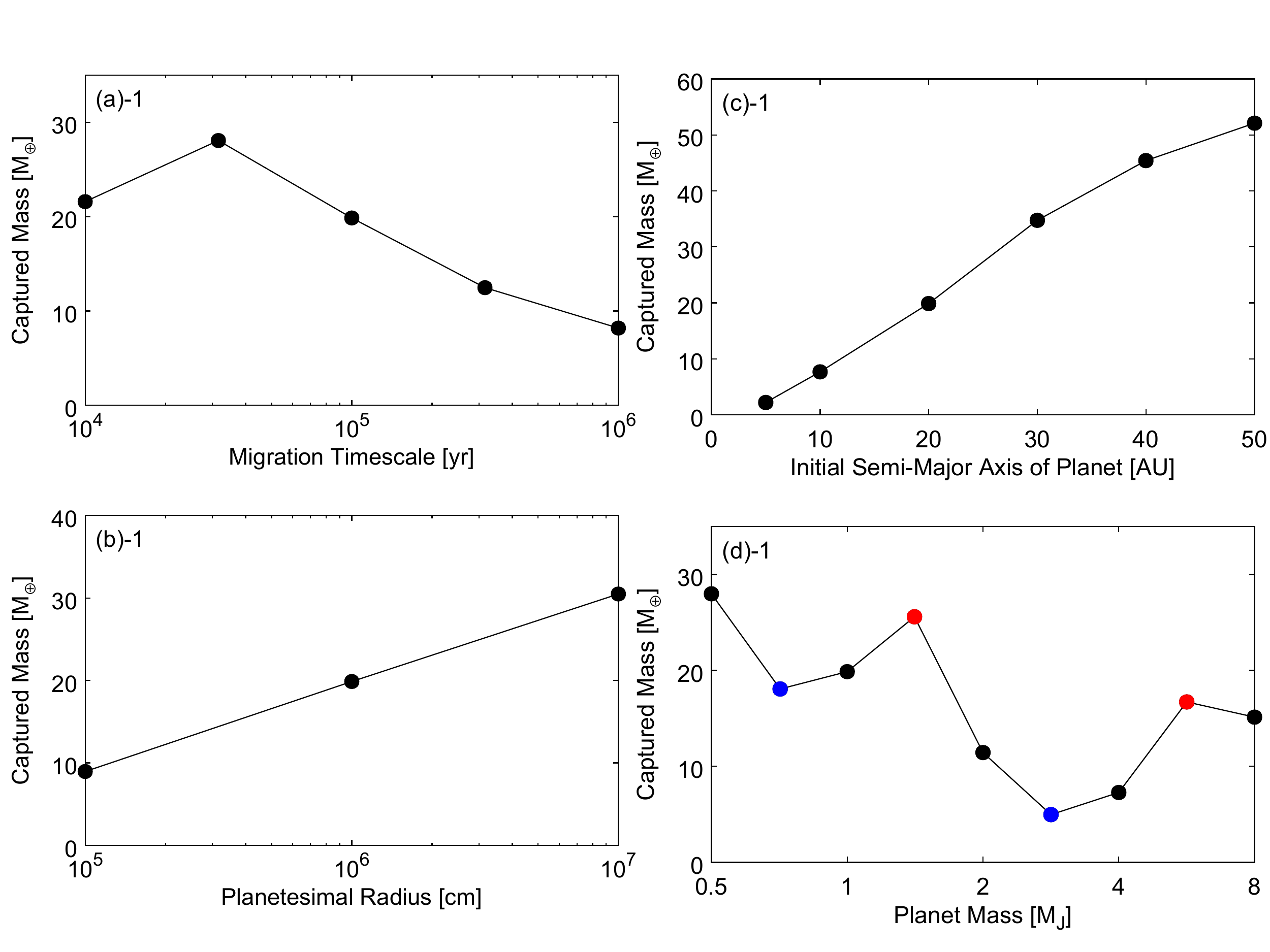}
\caption{\small Results of the parameter study of planetesimal capture by a migrating protoplanet performed by \cite{Shibata+2020}. Shown is the total mass of captured planetesimals in Earth masses as a function of: the migration timescale of the planet (a)-1, the radius of planetesimals (b)-1, the initial semi-major axis of the planet (c)-1, and the mass of the planet (d)-1.}\label{fig:enrichment-shibata}
\end{figure}

Many hot Jupiters are found to be highly enriched in heavy elements compared to their stellar-host metallicity. Theory suggests that some of these planets are expected to consist of several tens and even hundreds of Earth-masses of heavy elements \citep[although still with large uncertainties, see][]{Miller+2011,Thorngren+2016,wakeford+2017,sing+2018}. This introduces great challenges to the giant planet formation \textcolor{black}{theories} since the expected enrichment from the standard formation process is very moderate and typically cannot exceed about 20 $M_{\oplus}$ \cite[e.g.][]{Shibata+2019}. As a result, the enrichment of the planets must be explained.

One could imagine that the planetary enrichment occurs after the planet has formed and gas accretion terminated, a natural pathway being planetesimal capture. This process, however, is quite inefficient once the giant planets approach their final mass values \citep[being able to supply a few Earth-masses of heavy elements at most,][and references therein]{turrini+2015} and becomes even less efficient in presence of post-formation migration. \textcolor{black}{\cite{Shibata+2020} and \cite{turrini+2021} recently investigated the enrichment of warm gas giants via planetesimal capture during inward migration. \cite{Shibata+2020} performed orbital simulations of migrating giant planets of different masses and planetesimals in a protoplanetary gaseous disc and inferred the heavy-element mass that is accreted by the planet. \cite{turrini+2021} focused on Jovian planets but traced also their mass and radius evolution during their migration.}

\begin{tcolorbox}
\cite{Shibata+2020} found  that \textbf{migrating giant planets capture planetesimals with total masses of several tens of Earth masses}, if the planets start their formation at tens of au in relatively \textcolor{black}{compact discs (less than $100$ au, as generally assumed for the solar nebula)}. \textcolor{black}{A similar result is obtained in the study by \cite{turrini+2021} focusing on large (gas extending to a few $100$ au)} but not exceedingly massive discs (disc masses of the order of a few $10^{-2}$ M$_\odot$) if the giant planets have formation regions comparable to those recently observed by ALMA, i.e. extending from a few tens to about a hundred au from the host star \citep[e.g.][]{andrews+2018}.
\end{tcolorbox}

Depending on the characteristics of the considered protoplanetary disc, planetesimal capture seems to be efficient in a rather limited range of semi-major axis \citep{Shibata+2020} or for migration tracks spanning several tens of au \citep{turrini+2021}. Nevertheless, both studies showed that the total captured planetesimal mass increases with increasing  migration distances. It was also shown that mean motion resonances trapping and aerodynamic gas drag inhibit planetesimal capture of a migrating planet, and therefore large scale migration and/or massive/enriched discs are required to explain the enrichment of planets with several tens Earth masses of heavy elements. 

Figure \ref{fig:enrichment-shibata} summarizes the results of the study performed by \cite{Shibata+2020}, which suggests that enriched giant exoplanets at small orbits have not formed {\it in situ} since they must have migrated inward in order to accrete large amounts of heavy elements. \textcolor{black}{As will be discussed in more detail in Sect. \ref{sec:architectures}, however, recent population studies investigating the architectures of known multi-planets extrasolar systems \citep{zinzi+2017,laskar+2017,turrini+2020,he+2020} suggest} that a significant fraction of these planetary systems underwent or are crossing phases of chaotic evolution possibly associated to migration by planet-planet scattering \citep{rasio+1996,weidenschilling+1996}. 

A widespread presence of chaos-driven migration in the life of planetary systems, in alternative or in conjuction with disc-driven migration, would introduce a layer of uncertainty in unequivocally linking the formation and dynamical history to its heavy element enrichment. As an example, a giant planet forming and migrating between 30 and 20 au while embedded in the disc, as in the accretion tracks by \cite{Shibata+2020}, and later being scattered to a fraction of au by planet-planet scattering could be characterized by the same heavy element enrichment than a giant planet that started forming at about 10 au and experienced only disc-driven migration (see e.g. the top right panel of Fig. \ref{fig:enrichment-shibata}). As we will illustrate in Sect. \ref{sec:tracers}, however, the different compositional signatures of the accreted heavy elements allow for breaking this degeneracy.

\subsection{Envelope enrichment through core formation}\label{sec:envelope}

Several recent works follow the ablation of accreted solids in the gaseous envelope in the planet formation phase and find a significant pollution of the growing gas envelope by the accreted solids \citep[e.g.][]{brouwers+2018,valleta+2018,boden+2018}. It is found that when the planetary core mass is less than a few Earth masses most of the accreted solids, both rocks and ices, are deposited in the gaseous envelope and don’t reach the core. As the formation processes continues, the internal temperatures increase, the gravity becomes stronger and the gas is denser, and thus a larger fraction of the solids stays in the envelope.

\begin{tcolorbox}
The resulting structure right after the planet formation is then of a \textbf{gradual composition distribution instead of a core-envelope structure}. This primordial gradual metal distribution may evolve to a metal enrichment of the envelope in the long term. Under certain conditions, that are common in gas giant interiors, convection and therefore \textbf{convective-mixing can spread the composition gradient to the outer envelope in time}, and increase the outer envelope metallicity \citep{vazan+2015}.
\end{tcolorbox}

\begin{figure}[t]
\centering
\includegraphics[trim= 0.5cm 0cm 0.9cm 0.8cm, clip, width=0.5\columnwidth]{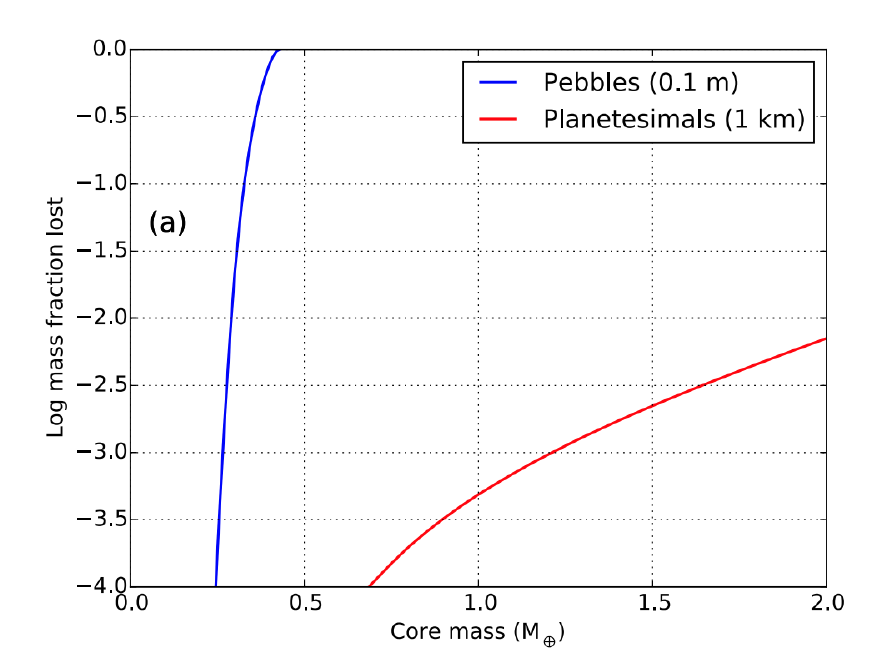}
\includegraphics[trim= 0cm 0cm 0.8cm 0.5cm, clip, width=0.475\columnwidth]{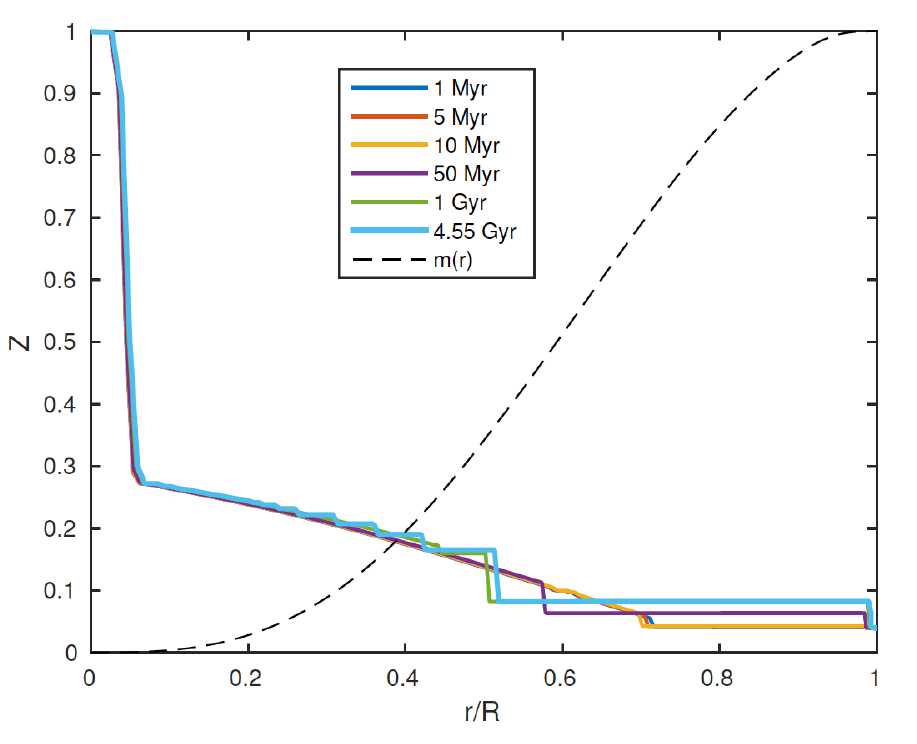}
\caption{{\it Left:} Mass deposition in the envelope (lost) as a function of the growing core mass, for rocky planetesimals (red) and pebbles (blue). Fragmentation is ignored. Rocky planetesimals lose approximately 1\% of their mass when the forming planet has a core mass of 2 Earth masses, whereas the pebbles are fully evaporated before the planet's core mass reaches 0.5 Earth mass \citep{brouwers+2018}. {\it Right:} the change in time in the metal distribution (Z) in the interior of Jupiter. At early ages convective-mixing is efficient in the outer shallow composition gradient, while the inner steeper gradient remains stable \cite{vazan+2018}.}
\label{fig:rock-vazan}
\end{figure}

The measured metal enrichment by the Ariel mission, and the derived statistics for different planetary types, can be used to constraint the formation outcomes and the convective behaviour. For example, in absence of fragmentation the envelope enrichment is greater in the case of pebble accretion than in planetesimal accretion. As is shown in the left panel of Fig.\ref{fig:rock-vazan}, pebbles dissolve better and earlier in the envelope than planetesimals, and therefore results in a more enriched envelope for the same planetary mass. The break-up of planetesimals during accretion can enhance the envelope pollution that is shown in Fig.\ref{fig:rock-vazan} \citep{mordasini+2015,valleta+2018}. Overall, ablation of pebbles is more efficient for small core/envelope masses while planetesimal break-up and ablation play a significant role for envelope masses greater than a few Earth masses \citep{mordasini+2015,podolak+2020}.

After the formation phase the local metal enrichment can be redistributed in the planet's envelope by convective-mixing. Long-term evolution of the structure by convective-mixing successfully explains the properties of our Solar System giant planets \citep{vazan+2016,vazan+2018,vazan+2020}, and is expected to take place in giant exoplanet interiors. The efficiency of the mixing depends on the metals distribution: an outer moderate enrichment tend to mix efficiently, while a deeper steeper distribution remains stable, as is shown in the right panel of Fig.\ref{fig:rock-vazan} for Jupiter. Thus, the initial metal enrichment by the formation building blocks affects the long-term atmospheric abundances of the planet.

\subsection{Compositional signatures of different formation regions}\label{sec:tracers}

\begin{figure}[t]
\centering
\includegraphics[trim= 1cm 0cm 1cm 1.5cm, clip, width=0.49\columnwidth]{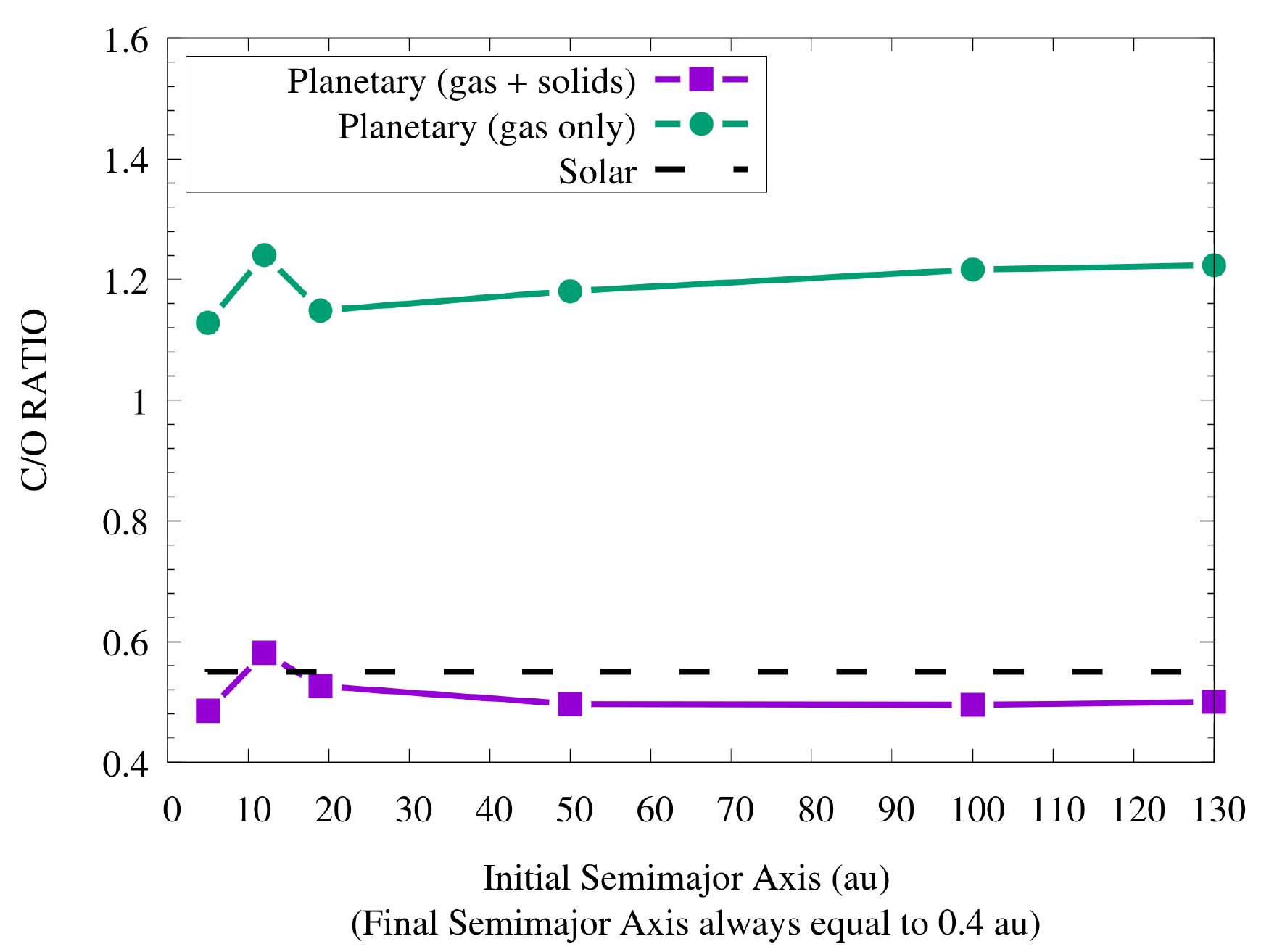}
\includegraphics[trim= 1cm 0cm 1cm 1.5cm, clip, width=0.49\columnwidth]{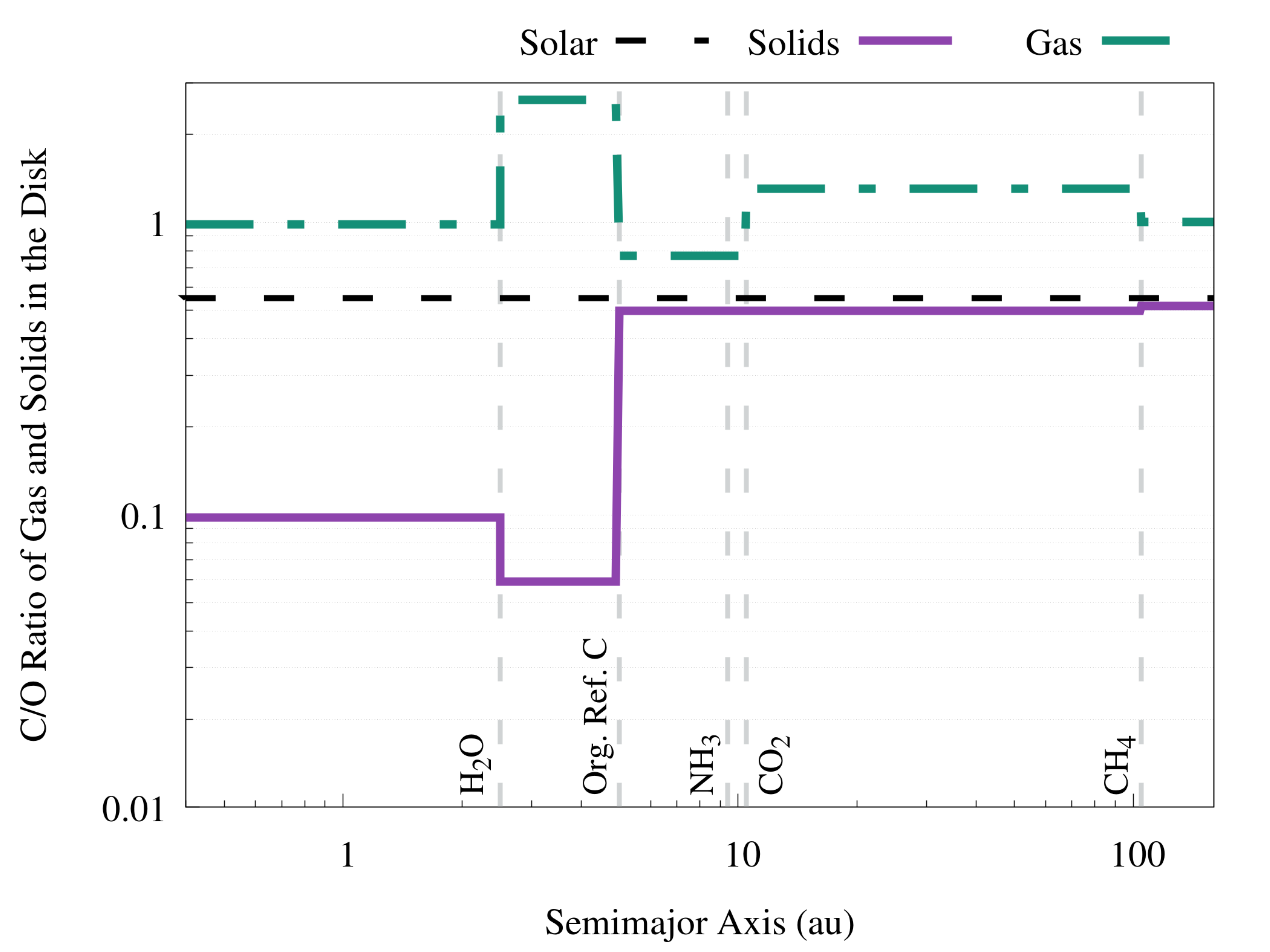}
\caption{\textit{Left }: C/O ratio of giant planets undergoing extensive migration from formation regions consistent with those observed with ALMA. \textit{Right }: C/O ratio of the gas and solids in the protoplanetary disk taking into account the roles of refractories and refractory organics as carriers of O and C (see main text and \citealt{turrini+2021} for details). The CO$_2$ snowline is located at about 10 au and the disc gas is dominated, from that point outward, by CO and CH$_4$ (C/O$\gtrsim$1). As a result, giant planets forming beyond the CO$_2$ snow line and accreting limited solids will have C/O$\gtrsim$1, while giant planets accreting large quantities of solids will have C/O slightly smaller than the stellar value. Figure adapted from \cite{turrini+2021}.}
\label{fig:C/O-turrini}
\end{figure}

\begin{figure}[t]
\centering
\includegraphics[width=0.99\textwidth]{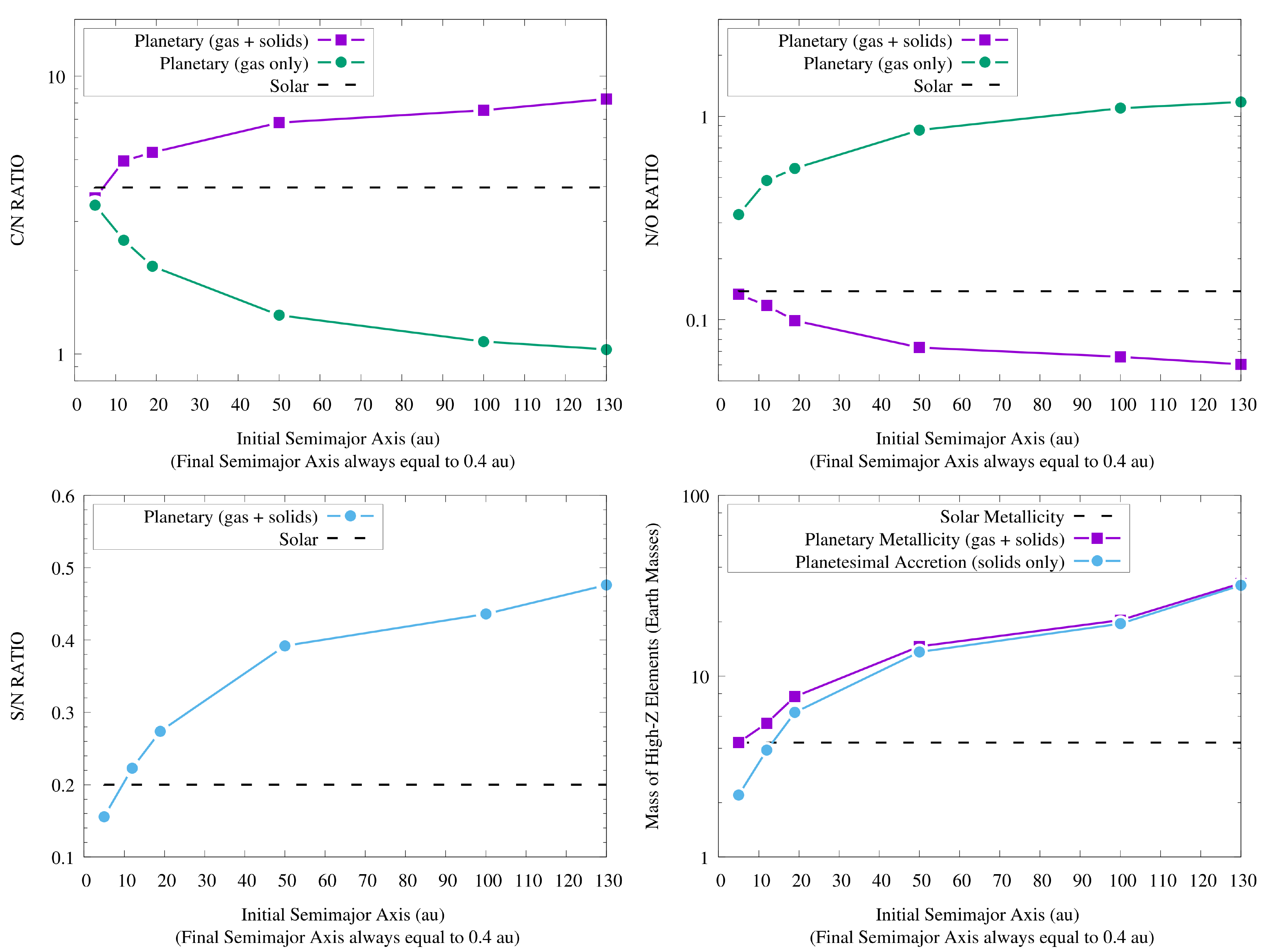}
\caption{C/N (top left), N/O (top right) and S/N (bottom left) elemental ratios of a Jupiter-sized giant planet as a function of different formation regions and migration tracks \citep{turrini+2021}. The Jovian planet starts its formation at the specified orbital distances and undergoes disc-driven migration until it becomes an hot Jupiter. The horizontal dashed lines indicate the stellar elemental ratios, assuming a solar composition for the host star and the protoplanetary disc \citep{palme+2014}. The different curves in the top half of the figure refer to different mass growth scenarios involving the accretion of both gas and planetesimals (``gas + solids'') or the sole gas (``gas only''). Also shown is the comparison between the total mass of heavy elements accreted by the giant planet and the one due only to planetesimal capture (bottom right), which highlights how the S/N ratio can be used as a proxy into the planetesimal contribution to metallicity. Figure from \cite{turrini+2021}.}
\label{fig:enrichment-turrini}
\end{figure}

\textcolor{black}{A number of studies have been devoted in recent years to link the information supplied by the composition of giant exoplanets to their formation and migration histories. Before discussing the insight they provide, however, it is important to point out that their task is made particularly challenging by our still incomplete understanding of the chemical environment of protoplanetary discs, the birthplace of giant planets. As highlighted by the discussions in Sects. \ref{sec:protoplanetary_discs} and \ref{sec:organics}, even if modern observational facilites are providing an unprecedented view of protoplanetary discs, their physical and compositional characterization is still hindered by a number of uncertain parameters and observational limitations. A particularly critical source of uncertainty is associated with the initial molecular setup of protoplanetary discs, as it is unclear whether protoplanetary discs inherit their composition from the prestellar phase or undergo a complete reset due to the radiation environment of their young stars.\\
\indent As discussed in Sect. \ref{sec:organics}, the results of the recent comparisons between the volatile inventory of the comets in the Solar System and Solar-like protostellar systems appear to support a strong role of inheritance from the prestellar phase \citep{drozdovskaya+2019,bianchi+2019,altwegg+2019}. However, even in this scenario, discs are expected to chemically evolve \citep{eistrup+2016,eistrup+2018} and cool down over time. Their temperature decrease will cause the snowlines of the more volatile elements to drift inward with respect to their original positions \citep{panic+2017,eistrup+2018}. Finally, the differential radial drift of the dust grains with respect to the gas will cause volatile elements initially frozen on the grains to cross the snowlines in the disc and sublimate, locally enriching the gas \citep{piso+2016,booth+2019b,bosman+2019}. A recent, detailed review of these processes and how they are expected to shape the compositional environment of protoplanetary discs is provided by \cite{oberg+2020}.\\
\indent It is important to note, however, that the magnitude of the previously listed effects is linked to the abundance of dust grains and pebbles in discs and that the conversion of dust and pebbles into planetesimals act to reduce the rate of compositional evolution of protoplanetary discs. The smaller surface-to-volume ratio of planetesimals with respect to dust and pebbles will slow down the rate of gas-grain chemistry and, due to the thermal inertia of the planetesimals that effectively isolate the ices trapped in their interior, will limit the effects of ice sublimation at the crossing of snowlines \citep[e.g.][]{turrini+2021}. Observational evidences from both protoplanetary discs \citep{manara+2018} and meteorites in the Solar System \citep{scott2007} points toward an efficient conversion of the bulk of dust into planetesimals on a timescale of less than 1 Myr, and possibly of the order of 10$^5$ years \citep{schiller+2018}. Such conversion would thus appear to proceed at a pace comparable or faster than the global evolution timescale (of the order of 1 Myr) estimated by astrochemical models of evolving discs \citep{eistrup+2018}, suggesting the possibility of an early ``freezing'' of the composition of protoplanetary disks.} 

\textcolor{black}{While we are still limited by our incomplete understanding of the compositional nature of protoplanetary discs, a growing literature has been focusing over the past decade on exploring} the link between the abundances of the two most abundant high-Z elements, carbon (C) and oxygen (O), and the planet formation process \cite[see e.g.][and references therein for recent overviews]{madhusudhan+2016,madhusudhan2019}. The general expectation since the early results from \cite{oberg+2011} is for low metallicity giant planets, where the bulk of C and O are accreted from the gas, to be characterized by super-solar C/O ratios, while for high-metallicity giant planets, where C and O are dominated by the capture of solids, to be characterized by sub-solar C/O ratios. \textcolor{black}{As a consequence, studies have been investigating the possibility to use the C/O ratio as a proxy into the formation region of giant planets (see e.g. \citealt{madhusudhan+2016} and references therein for an overview and \citealt{mordasini+2016}, \citealt{cridland+2019} for recent results).}

Critical factors to this end, however, are the distribution of C and O across the different phases (rocks, organics, and ices) and \textcolor{black}{volatile molecules (e.g. H$_{2}$O, CO$_{2}$, CO, CH$_{4}$), our understanding of which has been significantly evolving over the past decade thanks to the data provided by meteorites, comets, polluted white dwarfs, and protoplanetary discs  \citep[e.g.][]{lodders+2010a,lodders+2010b,oberg+2011,johnson+2012,palme+2014,thiabaud+2014,marboeuf+2014a,marboeuf+2014b,bergin+2015,mordasini+2016,bardyn+2017,isnard+2019,doyle+2019,cridland+2019,altwegg+2019,fulle+2019,rubin+2020,turrini+2021}, and the extension of the planet-forming region in discs, recently put into question by observational surveys of protoplanetary discs with ALMA \citep[e.g.][]{alma+2015,isella+2016,fedele+2017,fedele+2018,long+2018,andrews+2018}. \\
\indent A study performed in the framework of the Ariel Consortium (\citealt{turrini+2021}, see Fig. \ref{fig:C/O-turrini}, left plot) confirms the general picture described above for the planetary C/O ratio} also in the case of giant planets forming at tens and beyond one hundred of au from the host star, as suggested by the results of ALMA surveys. The same results, however, show how for giant planets forming so far from their host stars the information provided by the C/O ratio may be less detailed than expected. 
\textcolor{black}{In the framework of the inheritance scenario coupled with the early conversion of dust into planetesimals considered in the study, the reason for this is easily understood if one considers that, due to their higher volatility, CO and CH$_4$ condense about a order of magnitude farther aways from the star than CO$_2$ (see Fig. \ref{fig:C/O-turrini}, right plot). This in turn means that over a large fraction of the planet-hosting region suggested by ALMA's observations, the gas in the disc will be populated mainly by CO and CH$_4$. Giant planets forming beyond the CO$_2$ snowline will therefore accrete material from a region where the C/O of the gas will be dominated by the contributions of these two molecules (C/O$\gtrsim$1, see the right plot of Fig. \ref{fig:C/O-turrini}).} While gaseous CO and gaseous CH$_4$ will increase the C abundance of the gas and reduce that available for condensates, the abundances of these molecules will cause the C/O ratio of solids in this region to be only slightly smaller than the stellar value (see Fig. \ref{fig:C/O-turrini}, right plot). 

\textcolor{black}{As a result, giant planets starting their formation at orbital distances spanning the range revealed by ALMA surveys will accrete significant fractions of their mass, if not most of it, beyond the CO$_2$ snowline. Those giant planets whose mass growth is dominated by gas (low metallicity giant planets), e.g. forming across orbital regions previously depleted of planetesimals by the formation and migration of another giant planet, will have C/O$\gtrsim$1 \citep{turrini+2021}. Those capturing significant amounts of solids in the form of planetesimals (high metallicity giant planets) will have C/O slightly below than the stellar value almost independently on their exact formation region (see Fig. \ref{fig:C/O-turrini}, left plot, and \citealt{turrini+2021}). The limited changes in the C/O values shown in the left plot of Fig. \ref{fig:C/O-turrini} are smaller than the accuracy of current retrieval tools \citep[e.g.][]{barstow+2020}, meaning that those C/O values would be observationally indistinguishable from each other.} This translates in the fact that the C/O ratio might only allow to distinguish low metallicity, gas-dominated giant planets from high metallicity, solid-enriched giant planets and provide the information that they formed and captured most of their heavy elements farther out than the CO$_2$ snowline \citep{turrini+2021}. 

\textcolor{black}{The picture discussed above has been derived assuming the compositional inheritance of the volatile materials in the protoplanetary disc from the pre-stellar phase. As discussed at the beginning of this section and in Sect. \ref{sec:organics} (see also \citealt{altwegg+2019} and \citealt{oberg+2020} for more detailed discussion), while there are lines of evidence supporting such a scenario, it does not represent the only possible compositional setting for the planet formation process. The partitioning of the volatile molecules between gas and planetesimals can be markedly different in a compositional reset scenario, in principle making the picture described above for the planetary C/O ratio invalid. As discussed in \citet{turrini+2021}, future studies will need to quantify the effects of the different compositional scenarios on the planetary C/O ratio for giant planets forming over a wide range of orbital distances, to clarify the limits of its diagnostic power. Similarly, the effects of different couplings between the planet formation and the disc evolution timescales will need to be explored. Nevertheless, the available observational data on the roles of refractories and refractory organics as carriers of O \citep{lodders+2010a,lodders+2010b,palme+2014,doyle+2019} and C \citep{bergin+2015,bardyn+2017,isnard+2019} support the possibility that the quantitative changes in the planetary C/O ratio between one compositional scenario and another could be less marked than previously thought and, consequently, that the C/O ratio may provide only limited information.}

The vast coverage of Ariel in terms of molecules offers a straightforward way out of this limitation by allowing for the use of multiple elemental ratios \citep{tinetti+2018,turrini+2018}. \textcolor{black}{An illustrative example is provided by Fig. \ref{fig:enrichment-turrini}, which shows the results obtained in the study by \citep{turrini+2021}} using an extended set of four elemental ratios including, in addition to C and O, other cosmically abundant elements as nitrogen (N) and sulphur (S). The inclusion of N alongside C and O allows for computing two additional ratios: N/O and C/N. \textcolor{black}{Due to their higher volatility of N with respect to C and O, the N/O ratio} grows with migration for low metallicity giant planets and decreases for high metallicity ones, while C/N behaves the opposite way. The farther the giant planet starts its migration from the host star, the more its C/N and N/O ratios will diverge from the stellar ones \citep{turrini+2021}. 

The inclusion of S alongside N allows for computing the S/N ratio: given that the bulk of S is efficiently trapped into refractory solids  \citep[e.g.][]{lodders+2010a,lodders+2010b,palme+2014,kama+2019} while the bulk of N remains in gas phase as \textcolor{black}{highly volatile } N$_2$ for most of the extension of discs \citep[e.g.][]{pollack+1994,eistrup+2016,eistrup+2018,oberg+2019,bosman+2019}, this ratio offers a direct probe into the planetary metallicity \textcolor{black}{and, specifically, the fraction of the planetary metallicity due to the accretion of planetesimals (see Fig. \ref{fig:enrichment-turrini}). The S/N ratio, therefore, can be used to constrain, independently on the knowledge of the planetary mass and radius, the disc-driven migration experienced by the giant planet as discussed in Sect. \ref{sec:metallicity} (see \citealt{turrini+2021} for a discussion).} Recent works focusing on the study of Jupiter's formation in the Solar System \citep{oberg+2019,bosman+2019} further highlighted how the combination between a super-stellar metallicity (e.g. obtained through the mass-radius relationship) with a stellar S/N ratio in a giant planet can indicate its formation beyond the N$_{2}$ snowline (N$_{2}$ being the main N carrier in protoplanetary discs). \textcolor{black}{Note that, due to its high volatility, N$_2$ condenses as ice at a few tens of au even in cold discs \citep{eistrup+2016,eistrup+2018, oberg+2019}, while for warmer discs (e.g. 280 K at 1 au, as generally assumed for the solar nebula and adopted by \citealt{turrini+2021}) N$_2$ may remain in gas form until a few hundreds of au from the star, farther out than even the planet-hosting region suggested by ALMA's surveys.}


\begin{tcolorbox}
The use of \textbf{multiple elemental ratios involving elements of different volatility} permitted by Ariel's spectral coverage allows to greatly \textbf{reduce the degeneracy intrinsic in any single measure} and to more robustly constrain the formation and dynamical history of giant planets \citep{turrini+2021}.
\end{tcolorbox}

\begin{figure}[t]
\centering
\includegraphics[width=\textwidth]{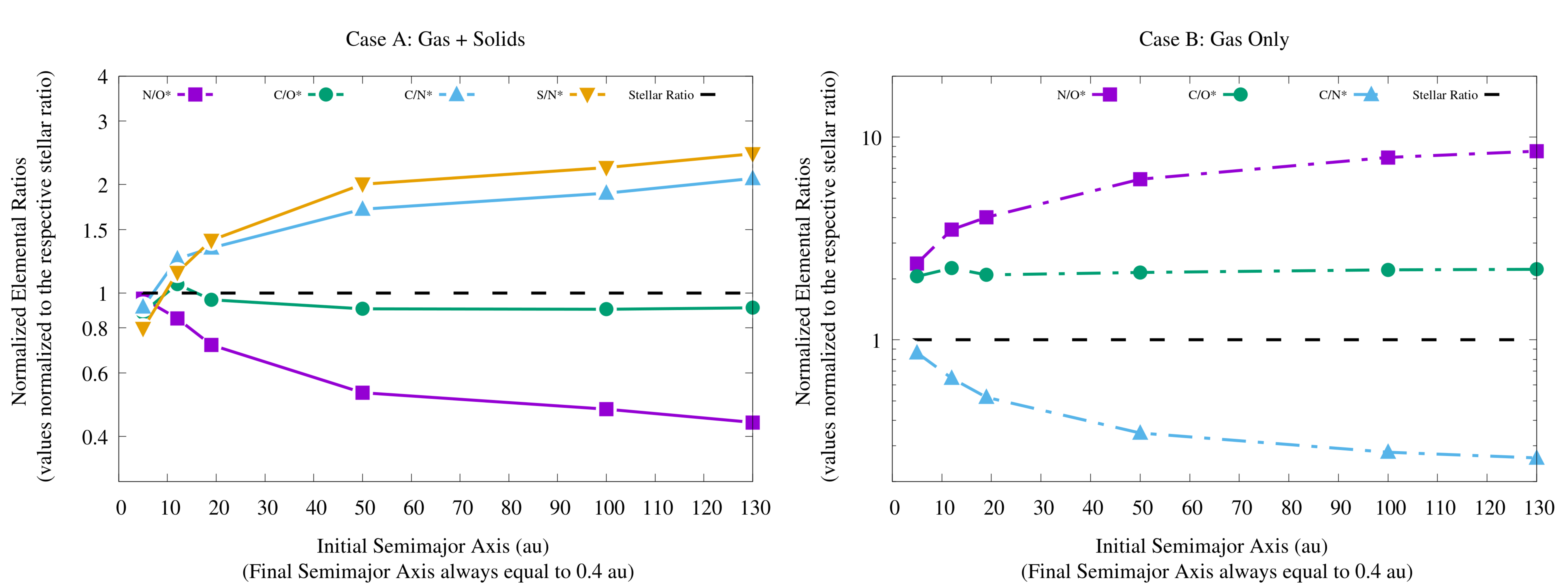}
\caption{\emph{Left:} comparison of the normalized C/O, N/O, C/N, and S/N elemental ratios of the gaseous envelope when the metallicity of the giant planet is dominated by the accretion of planetesimals (high metallicity case). \emph{Right:} comparison of the normalized elemental ratios in the gaseous envelope when the metallicity of the giant planet is dominated instead by the accretion of gas (low metallicity case). Each elemental ratio is normalized to the relevant stellar elemental ratio. Figure from \cite{turrini+2021}.}\label{fig:normalized_ratios}
\end{figure}

\textcolor{black}{
It is important to point out that the discussion above focuses on specific absolute values that have been derived assuming a composition of the protoplanetary disc matching the protosolar composition \citep[see e.g.][and references therein]{asplund2009,lodders+2019}. As discussed in Sect. \ref{sec:stellar_characterizarion}, different stars will be characterized by different metallicities and, more importantly, different elemental ratios. As a result, the elemental ratios of planets orbiting different stars cannot be directly compared and the specific values reported above (e.g. C/O $>$ 1) should not be considered as absolute references.\\
\indent This obstacle can be overcome with the use of planetary elemental ratios normalised to their relevant stellar values, analogously to the case of the normalized metallicities values adopted by \cite{Thorngren+2016}. The use of normalized elemental ratios (not necessarily limited to the cases of C/N, N/O, C/O and S/N discussed above) removes the intrinsic compositional variability between different planetary systems and opens up the possibility of more reliable comparisons between the respective formation and migration histories of giant planets orbiting different stars.\\
\indent Furthermore, as discussed by \cite{turrini+2021} the use of normalized elemental ratios associated with elements characterized by different volatility provides additional constraint on the nature of giant planets. The C/O, C/N, N/O and S/N ratios normalised to their stellar values (indicated with the superscript *) reveal that high metallicity giant planets will be characterized by C/N* $>$ C/O* $>$ N/O* (see Fig. \ref{fig:normalized_ratios}). Gas-dominated, low metallicity giant planets, instead, will be characterized by N/O* $>$ C/O* $>$ C/N* (see Fig. \ref{fig:normalized_ratios}). Giant planets for which planetesimal accretion is the main source of metallicity will have S/N* $>$ C/N*, while those for which both gas and solids contribute to the metallicity will have instead C/N* $>$ S/N* (see Fig. \ref{fig:normalized_ratios}).\\
\indent Finally, since the normalization to the stellar values brings the planetary elemental ratios of elements with different cosmic abundances on a common scale, any element whose main carrier is characterized by a lower or similar volatility than S \citep[see e.g.][]{palme+2014,turrini+2018} can be used to compute normalized elemental ratios with respect to N and gain insight on the source of the planetary metallicity \citep{turrini+2021}. The use of normalized elemental ratios therefore allows to compare the constraint on the metallicity derived for different giant planets using different low-volatility elements (e.g. S/N*,Al/N*,Na/N*,Cr/N*).
}

\section{High-density Planets: Formation and Atmospheres}\label{sec:high_density_planets}

The \textit{Kepler} exoplanet survey revealed that a vast majority of close-in exoplanets are smaller in size than Neptune \citep{batalha+2014}. Such planets are called \textit{high-density planets} in this manuscript, as the bulk of their mass is represented by condensates with higher densities that the gas providing most of the mass of gas giants like Jupiter and Saturn. Given their high occurrence, understanding their formation is a central issue in exoplanetary science. 

High-density planets, in general, are formed in a complicated way through various processes including solid and gas accretion, orbital migration, giant collisions, late veneers, mass loss, etc. Thus, the sole knowledge of basic physical properties such as mass, radius, and orbital elements is not enough to unveil their nature \citep[e.g.][and references therein]{tinetti+2018,turrini+2018}. The characterisation of their bulk and atmospheric compositions is therefore the key to understand the formation and diversity of low-gravity planets \citep{tinetti+2018,turrini+2018}.

\begin{figure}[t]
    \centering
    \includegraphics[trim= 1cm 0cm 0.9cm 0.8cm, clip,width=0.48\columnwidth]{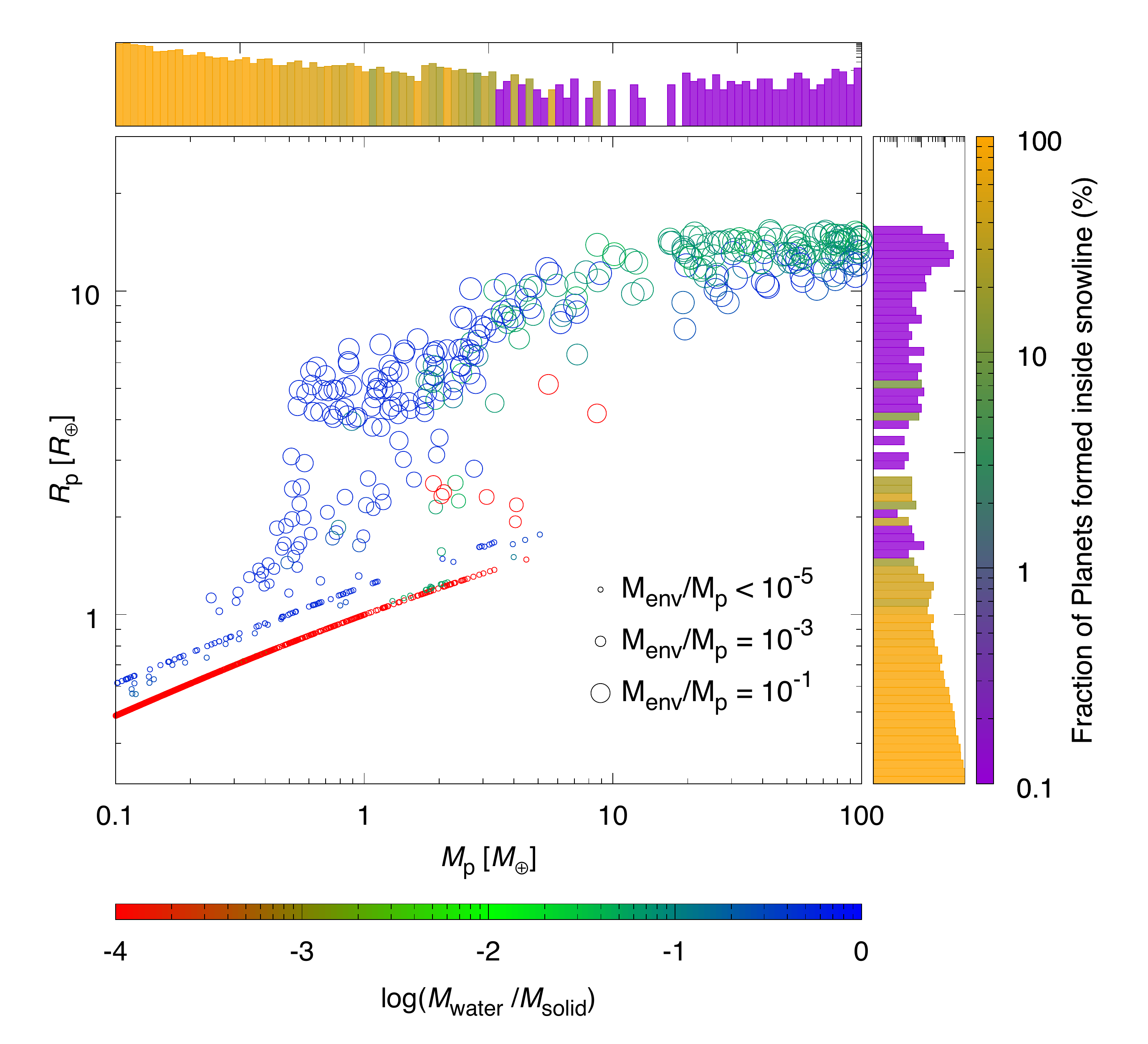}
    \includegraphics[trim= 1cm 0cm 0.9cm 0.8cm, clip,width=0.48\columnwidth]{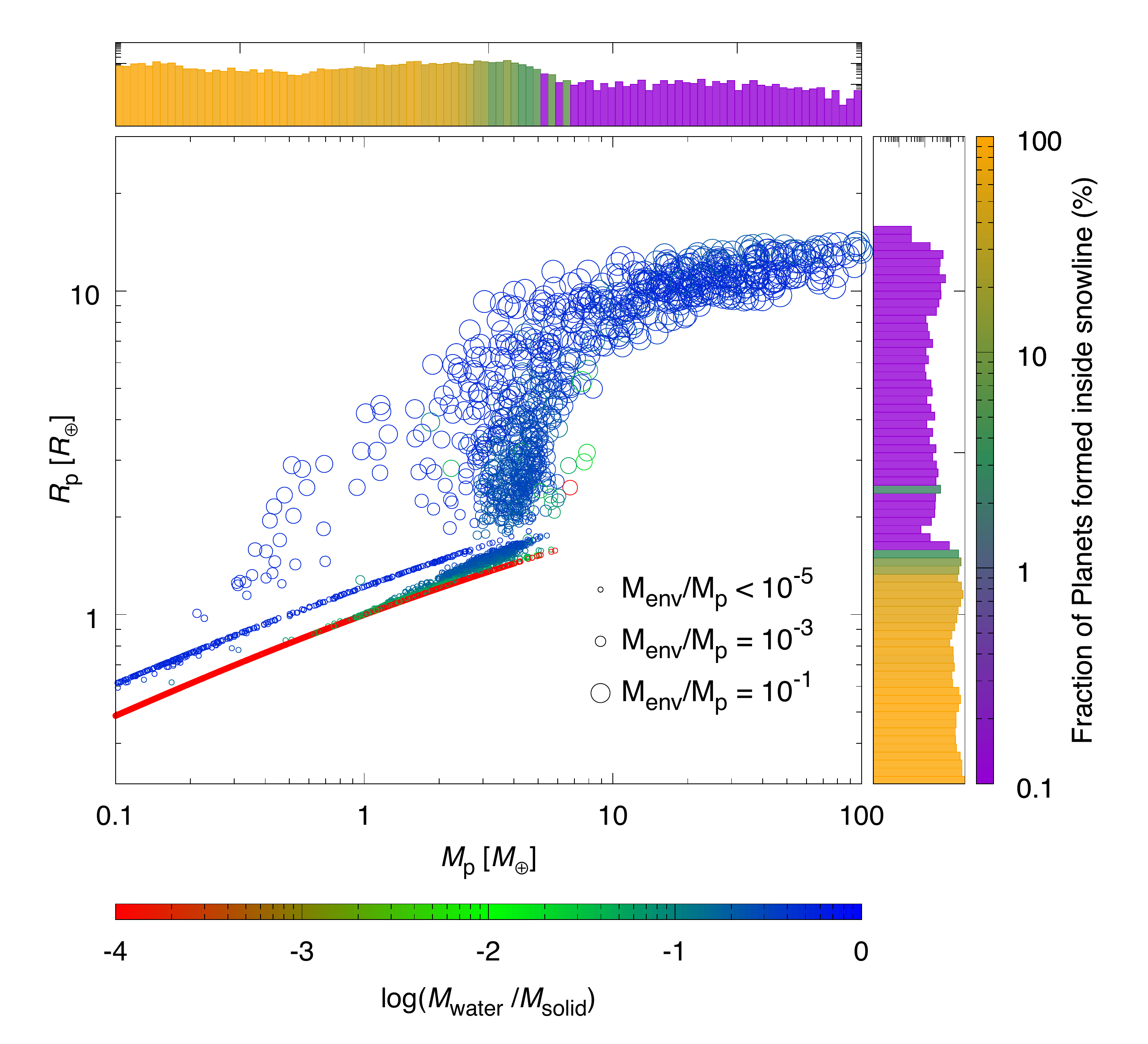}
    \caption{\small
    Theoretical prediction of masses, radii, and volatile contents of planets around a 0.3~$M_\odot$ star. 
    The population synthesis models include planetesimal accretion, gas accretion, orbital migration, collision of planetary embryos, viscous dissipation of the protoplanetary disc, and photo-evaporation of planetary atmospheres \citep[][Kimura et al. in prep.]{miguel+2019}. 
    The rate of the type-I migration differs between the two panels: 
    In the left and right panels, 1\% (slow) and 10\% (fast), respectively, of the migration rate from \cite{tanaka+2002} are assumed. 
    The synthesised planets are composed of a solid body (ice plus rock) and a H-He atmosphere. 
    In the mass-radius relationship diagram, symbols are colour-coded according to the total mass of water, which comes from icy planetesimals, and are sized according to the atmospheric mass relative to the solid planet mass. 
    Also, in the mass and radius histograms, where the number is given in log, the colour coding indicates the percentage of planets formed inside the snowline. 
    The synthesised planets have been sampled according to their transit probabilities.}
    \label{fig:pop_syn}
\end{figure}

One of the biggest uncertainties in the planet formation process is the orbital migration that occurs via angular momentum exchange between the planet and the circumstellar disc (the so-called type-I migration). 
Planetary migration leads to the delivery of cold materials from beyond the snowline to the inner regions of discs and, thus, brings about a variety in composition of close-in planets. 
Since planetary migration occurs in a circumstellar disc composed predominantly of hydrogen and helium, migrating planets generally capture the surrounding disc gas by gravity to form an atmosphere. 
Such atmospheres of high-density planets are often termed primordial atmospheres or captured atmospheres. 

Figure~\ref{fig:pop_syn} shows the predicted masses, radii, and volatile contents of synthesised planets around M dwarfs of 0.3~$M_\odot$ with slow (\textit{left panel}) and fast (\textit{right panel}) migration. 
\textcolor{black}{Here we have carried out those calculations by adding the effects of atmospheric accumulation and loss \citep{kimura+2020} in the population synthesis models \citep{miguel+2019}}.
The symbols for radii of 1--4~$R_\oplus$ are shown with different colours and sizes, indicating that the high-density planets are diverse in bulk composition; namely, they have different ice-to-rock ratios and different atmospheric masses. 
As seen in the histograms, the bulk composition of high-density planets also differs depending on the migration rate. Thus, knowledge of bulk composition places a crucial constraint to migration rates. It is noticed, however, that some of the planets in Figure \ref{fig:pop_syn} have the same mass-radius relationships but different composition. Such degeneracy in composition prevents us from constraining the bulk composition \citep[see also][for a discussion]{turrini+2018}. Observation of their atmospheres with Ariel is of obvious significance.

While the disc gas consists predominantly of hydrogen, the atmospheres are not always hydrogen-dominated. 
Instead, they would contain heavier molecules than H$_2$ and He. Such contamination (or enrichment) occurs because of degassing from volatile-rich planetesimals and magma oceans \cite[e.g.][]{Nikolaou+2019} and chemical interaction between the atmospheric gas and minerals from magma oceans \citep[e.g.][]{kimura+2020}. In some extreme cases, the planets might lose all their primary atmosphere due to evaporation processes and interaction with the host star, and might have an outgassed, secondary atmosphere \citep{miguel+2011,Nikolaou+2019}. 

A mixture of both acquired and core-degassed volatiles is likely to form the atmospheric inventory. Moreover, the specifics of how volatile species chemically bond with rocky interiors found in solid (silicate mantle) or molten (magma ocean) state suggest that the sources of less soluble versus soluble species may differ. That is, CO and CO$_2$ that are less soluble in silicate melts could be provided directly from the captured disc gas, while H$_2$O could be provided from thermal evolution of the interior \citep{Nikolaou+2019} as well as upper atmosphere chemistry \citep{Ikoma+2006}. Thus, detailed investigation of atmospheric constituents helps us understand such processes, including contamination by and partitioning processes of heavy volatiles. 

\begin{figure}[t]
    \centering
    \includegraphics[width=0.7\columnwidth]{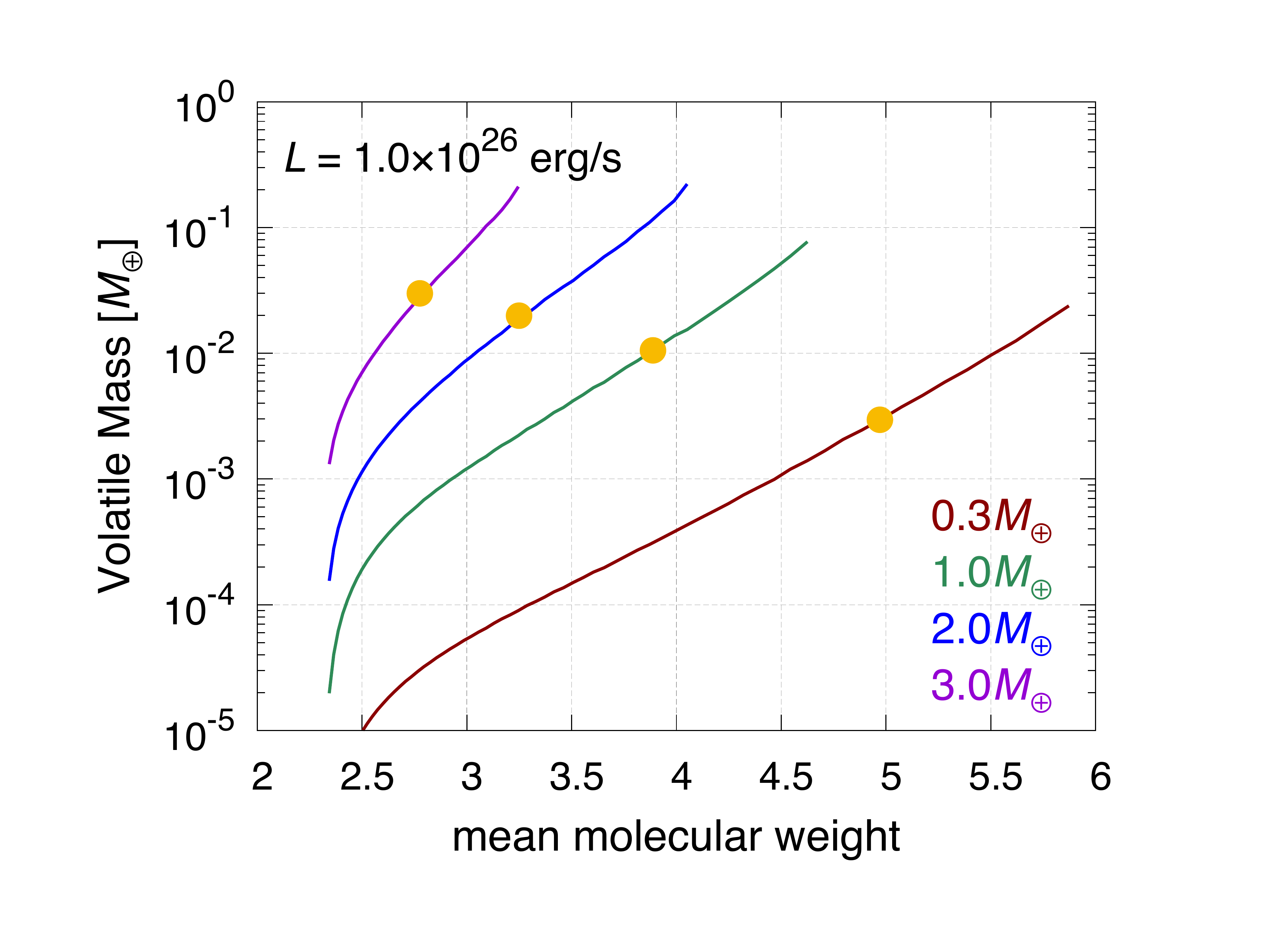}
    \caption{\small 
    Possible range of mean molecular weight of the atmospheric gas for sub- and super-Earths.
    We have calculated the mass of the enriched atmosphere in dynamical equilibrium with the protoplanetary disc as a function of the mean molecular weight by the same method as \cite{Ikoma+2018} and \cite{kimura+2020}.
    The orange symbols indicate the maximum amounts of volatile that would be degassed from a magma ocean with water content of 1~\%. 
    In these calculations we have assumed that the planet is located at 0.2~AU from an M dwarf of 0.3~$M_\odot$ and the energy flux in the atmosphere is $1 \times 10^{26} \, \mathrm{erg/s}$.}
    \label{fig:mean_molecular_weight}
\end{figure}

Contamination of heavy elements, however, tends to reduce the atmospheric scale height due to increase in mean molecular weight ($\mu$), and thereby hinders atmospheric characterisation via transmission spectroscopy. Figure~\ref{fig:mean_molecular_weight} shows the relationship between the total mass of H$_2$O contained in the atmosphere and the mean molecular weight of the atmospheric gas for several choices of the solid planet mass by the same method as \cite{kimura+2020}. Here we have calculated the structure and mass of the atmosphere enriched with water that is connected to the circumstellar gas disc. For reference, the orange symbols indicate the maximum amounts of volatiles that can be degassed from a magma ocean with H$_2$O content of 1~\%. 

As shown in this figure, the mean molecular weight of the atmosphere is at most five, which is about twice as high as that of the atmospheric gas with solar abundances. 
Note that we have ignored any carbon-based molecules such as CO$_2$ here for simplicity; for the gas of $\mu$ = 5, for example, if H$_2$O is replaced completely with CO$_2$, $\mu$ increases (and, thus, the pressure scale height decreases) by 10~\%.  

\begin{tcolorbox}
Ariel has the capability to \textbf{constrain the atmospheric mean molecular weight of high density planets} \citep{edwards+2019}. This can be achieved already with Tier 1 resolution for the most favourable cases. In the less favourable cases, additional observational time will be required to constrain the mean molecular weight, though this additional time is estimated to be less than what would be needed to achieve full Tier 2 resolution \citep{edwards+2019}.
\end{tcolorbox}

More refined assessments are ongoing (Mugnai et al., in preparation), with a particular focus on verifying the possibility of coupling the estimation of the atmospheric mean molecular weight with the detection of the main molecular constituents (especially water, \citealt{turrini+2018}). Nevertheless, the current picture indicates that Ariel should be able to provide indications on the \textit{primary/secondary atmospheres ratio among high-density planets} \citep{turrini+2018,edwards+2019}.

\begin{figure}[t]
\centering
\includegraphics[width=0.75\textwidth]{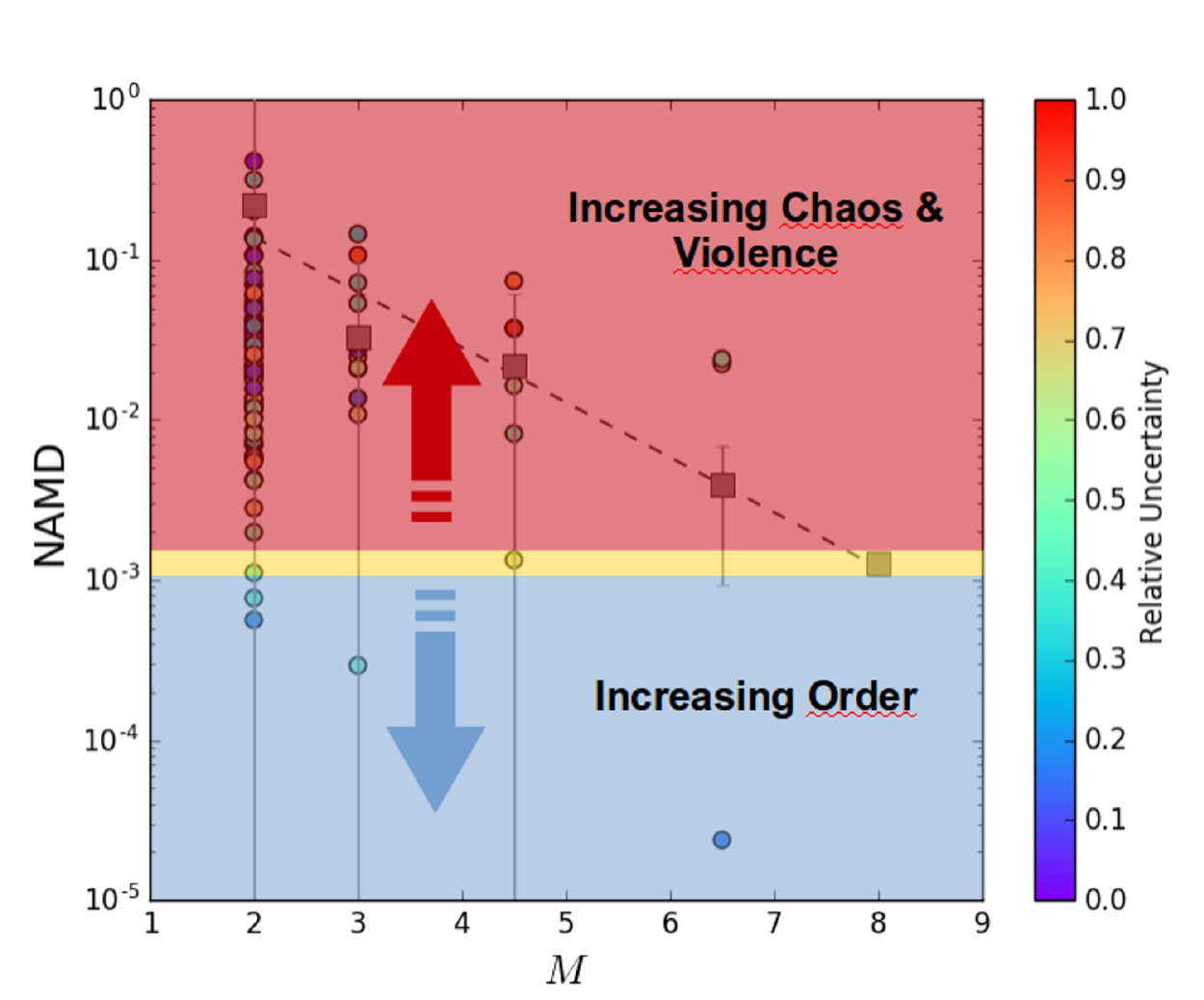}
\caption{Illustrative example of ``dynamical temperature'' scale built using Trappist-1 and the Solar System as reference systems. The underlying plot shows the dynamical excitation, quantified by the NAMD, of the 99 best-characterized planetary systems (filled circles) grouped according to the multiplicity of the planetary systems (i.e. their number of planets) with the systems with $M$=4 and $M$=5 and with $M$=6 and $M$=7 respectively grouped together to increase the statistics. Each planetary system is color-coded according to the relative uncertainty of its NAMD value. Also shown are the mean NAMD values of each multiplicity population (filled squares), computed as weighted-averages over the uncertainties of the individual systems. The overlaid coloured areas showcase an illustrative division between increasing likelyness of dynamical violence (red) and orderly evolution (blue), separated by an uncertainty region (yellow) centered on the Solar System. Figure adapted from \cite{turrini+2020}.}
\label{fig:dynamical_temperature}
\end{figure}

\textcolor{black}{
\section{Planetary Architectures: Dynamical Context to Composition}\label{sec:architectures}
Before moving to the conclusions it is worth emphasizing once again that, as discussed in Sect. \ref{sec:metallicity}, disc-driven migration is not the only dynamical process capable of delivering giant planets from their formation regions to the orbital distances where Ariel will observe them today. Other migration mechanims (planet-planet scattering, ejection from resonances, orbital chaos)
can achieve the same outcome while having markedly different implications for the composition of the  planets they affect \citep[see e.g.][and references therein]{turrini+2015,turrini+2018}. Furthermore, as discussed in Sect. \ref{sec:galactic_environment} there is emerging evidence suggesting a role played by the galactic environment in shaping the characteristics of planetary systems.\\
\indent Recent population studies of multi-planet systems highlight how their architectures record a strong role of violent processes, such as chaos and planet-planet scattering, in shaping the dynamical histories of known exoplanets \citep{limbach+2015,zinzi+2017,laskar+2017,turrini+2020,he+2020,bach-moller+2021}. As such mechanisms act when most solid mass in planetary systems has been incorporated into a limited number of massive bodies, the migrating planets they produce will encounter and accrete less material than their counterparts migrating in protoplanetary discs. At the same time, however, stochastic encounters between planets may result in catastrophic collisions with major implications for the composition and interior structure of the emerging planet.\\
\begin{tcolorbox}
\indent Consequently, this strong \textbf{role played by migration mechanisms other than disc-driven migration introduces a layer of uncertainty} in linking Ariel's compositional data to the formation histories of the observed exoplanets. The same population studies, however, suggest that the combined use of \textbf{metrics linked to the angular momentum of planetary systems} \citep{zinzi+2017,laskar+2017,turrini+2020} allows for extracting information from their architectures and \textbf{constrain their dynamical past}. 
\end{tcolorbox}
In particular, \cite{turrini+2020} and \cite{carleo+2021} have shown how the information provided by the normalized angular momentum deficit (NAMD), an architecture-agnostic measure of the dynamical excitation of planetary systems, allows to build a relative scale of violence of their past histories. Intuitively, the NAMD can be interpreted as the ``dynamical temperature'' of planetary systems: the higher the value, the more excited is the dynamical state of the system. As in the case of temperature, if one can identity meaningful reference values (as with the freezing and boiling points of water), it is possible to build a scale of dynamical excitation on which to measure the violence of the past of planetary systems.\\
\indent As discussed by \cite{turrini+2020} and \cite{carleo+2021}, Trappist-1 and the Solar System provide two such reference values, the first as a system characterized by an orderly and stable evolution \citep{tamayo+2017,papaloizou+2018} while the second as the boundary between orderly and chaotic evolution \citep{nesvorny2018}. As shown in Fig. \ref{fig:dynamical_temperature} the higher the NAMD of a planetary system with respect to that of the Solar System, the higher the likeliness that chaos and violent dynamical events sculpted its past.
Conversely, NAMD values increasingly closer to that of Trappist-1 are associated to increasing likeliness of stable and orderly histories.\\
\indent As a consequence, the measure of the ``dynamical temperature'' of planetary systems permitted by the NAMD can provide a dynamical context for the interpretation of Ariel's compositional observations. In other words, the \textit{combination of Ariel's observations} with the information provided by \textit{planetary system architecture} (specifically masses and orbital elements of its planets) will allow to extract additional and more detailed information on the history of the planets and their host system. It should be noted that the planetary physical and dynamical parameters don't need to be known at the time of Ariel's observations but can be included in the interpretation of Ariel's data at a later time, meaning that \textit{Ariel's scientific impact will grow over time}
}

\section{Concluding Remarks}

As introduced in Sect. \ref{sec:introduction} and further discussed in Sects. \ref{sec:giant_planets}, \ref{sec:high_density_planets}, and \ref{sec:architectures}, the planet formation process plays a fundamental role in shaping the final composition of planets and, consequently, of their atmospheres. Ariel's observations will therefore provide an unprecedented wealth of data to advance our understanding of planet formation in our Galaxy. However, as the discussion in Sects. \ref{sec:protoplanetary_discs}-\ref{sec:organics} highlights, a number of \textit{environmental factors} linked to the star and its own formation process \textit{affect the final outcome}: the galactic environment in which the star formation process takes places, the stellar composition and the thermal and physical structure and evolution of the protoplanetary disc.

As the implications of these environmental factors are still poorly constrained or understood, they can act as a \textit{source of uncertainty} or noise in the interpretation of the atmospheric data Ariel will provide and the reconstruction of the formation and evolution history of the observed planets. As a consequence, care should be taken, particularly during the initial phases of the nominal mission, to \textit{keep these factors into account in the selection of Ariel's targets} (and their stellar hosts) to \textit{minimize the free parameters} in this already complex problem.

The same considerations expressed above, however, also mean that the \textit{potential impact of Ariel's observations for understanding and quantifying the role played by these environmental factors is huge}, particularly when considering an extended mission and the even larger and more diverse observational sample it will bring. As illustrated in particular in Sect. \ref{sec:giant_planets}, the wide spectral coverage and the resulting large number of molecules that can be traced by Ariel means that the mission is uniquely suited to explore in unprecedented details and from different angles the \textit{link between the star formation and the planet formation processes}.


\section*{Acknowledgments}

D.T., S.F., S.M., E.S., and A.N. acknowledge the support of the Italian Space Agency (ASI) through the ASI-INAF contract 2018-22-HH.0. D.T., C.C., D.F., and L.P. acknowledge the support of the PRIN-INAF 2016 ``The Cradle of Life - GENESIS-SKA (General Conditions in Early Planetary Systems for the rise of life with SKA''. D.T., S.F., S.M. D.F, J.M., F.O., P.W. acknowledge the support of the Italian National Institute of Astrophysics (INAF) through the INAF \emph{Main Stream} project ``Ariel and the astrochemical link between circumstellar discs and planets'' (CUP: C54I19000700005). S.M. acknowledges support from the European Research Council via  the Horizon 2020 Framework Programme ERC Synergy ``ECOGAL'' Project GA-855130. M.K. acknowledges funding by the University of Tartu ASTRA project 2014-2020.4.01.16-0029 KOMEET ``Benefits for Estonian Society from Space Research and Application’’, financed by the EU European Regional Development Fund. J.M.D.K.\ gratefully acknowledges funding from the Deutsche Forschungsgemeinschaft (DFG, German Research Foundation) through an Emmy Noether Research Group (grant number KR4801/1-1) and the DFG Sachbeihilfe (grant number KR4801/2-1), as well as from the European Research Council (ERC) under the European Union's Horizon 2020 research and innovation programme via the ERC Starting Grant MUSTANG (grant agreement number 714907). The research of O.P. is funded by the Royal Society, through Royal Society Dorothy Hodgkin Fellowship DH140243. M.P. thanks the support to NuGrid from STFC (through the University of Hull's Consolidated Grant ST/R000840/1), and access to {\sc viper}, the University of Hull High Performance Computing Facility. M.P. acknowledges the support from the "Lendulet-2014" Program of the Hungarian Academy of Sciences (Hungary), from the ERC Consolidator Grant (Hungary) funding scheme (Project RADIOSTAR, G.A. n. 724560), by the National Science Foundation (NSF, USA) under grant No. PHY-1430152 (JINA Center for the Evolution of the Elements). M.P. also thanks the UK network BRIDGCE and the ChETEC COST Action (CA16117), supported by COST (European Cooperation in Science and Technology). M.I. thanks the support by JSPS KAKENHI 18H05439. C.D. acknowledges financial support from the State Agency for Research of the Spanish MCIU through the \textquotedblleft Center of Excellence Severo Ochoa\textquotedblright\ award to the Instituto de Astrof\'isica de Andaluc\'ia (SEV-2017-0709), and the Group project Ref. PID2019-110689RB-I00/AEI/10.13039/501100011033.

\bibliographystyle{spbasic}
\bibliography{bibliography.bib}

\end{document}